\documentclass[11pt]{article}

\usepackage{epsfig}
\usepackage{amsfonts,amssymb,amsmath}
\usepackage{vmargin}
\usepackage{bbold}

\numberwithin{equation}{section}

\newcommand{\D}{\ensuremath{{\cal D}}}

\newcommand{\G}{\ensuremath{{\cal G}}}

\newcommand{\N}{\ensuremath{{\cal N}}}

\newcommand{\J}{\ensuremath{{\cal J}}}
\newcommand{\I}{\ensuremath{{\cal I}}}

\newcommand{\ra}{\ensuremath{\rightarrow}}
\newcommand{\half}{\ensuremath{\frac{1}{2}}}

\newcommand{\be}{\begin{equation}}
\newcommand{\ee}{\end{equation}}

\newcommand{\ba}{\begin{eqnarray}}
\newcommand{\ea}{\end{eqnarray}}

\newcommand{\ns}{\normalsize}

\newcommand{\nn}{\nonumber}
\newcommand{\w}{\wedge}
\newcommand{\im}{\mathrm{Im\;}}
\newcommand{\re}{\mathrm{Re\;}}
\newcommand{\one}{\mathbb{1}}

\begin{document}

\begin{titlepage}

\title{
   \hfill{\ns hep-th/0701173\\}
   \vskip 2cm
   {\Large\bf Towards
Minkowski Vacua in Type II String Compactifications}
\\[0.5cm]}
   \setcounter{footnote}{0}
\author{
{\ns\large 
  \setcounter{footnote}{1}
  Andrei Micu$^1$\footnote{email: amicu@th.physik.uni-bonn.de}
  $^{,}$\footnote{On leave from IFIN-HH Bucharest}~,\;\;\; 
  Eran Palti$^2$\footnote{email: palti@thphys.ox.ac.uk}~,\;\;\; 
  Gianmassimo Tasinato$^{2}$\footnote{email: tasinato@thphys.ox.ac.uk}}
\\[0.5cm]
   $^1${\it\ns Physikalisches Institut der Universit\"at Bonn} \\
        {\it\ns Nussallee 12, D-53115, Bonn, Germany} \\[0.2em] 
   $^2${\it\ns Rudulf Peierls centre for Theoretical Physics, University of Oxford}\\
   {\it\ns Keble Road, Oxford, UK. } \\[0.2em] }

\date{}

\maketitle

\begin{abstract}\noindent
  We study the vacuum structure of compactifications of type II string
  theories on orientifolds with $SU(3)\times SU(3)$ structure. We argue that
  generalised geometry enables us to treat these non-geometric
  compactifications using a supergravity analysis in a way very similar to
  geometric compactifications. We find supersymmetric Minkowski vacua with
  all the moduli stabilised at weak string coupling and all the tadpole
  conditions satisfied. Generically the value of the moduli fields in the
  vacuum is parametrically controlled and can be taken to arbitrarily large
  values.
\end{abstract}

\thispagestyle{empty}

\end{titlepage}

%%%%%%%%%%%%%%%%%%%%%%%%%%%%%%%%%%%%%%%%%%%%%%%%%%%%%%%%%%%%%%%%%%%%%%%%%%%%%%%%%%%%%%%%%%%%%%%%%%%%%%%%%
\section{Introduction}
%%%%%%%%%%%%%%%%%%%%%%%%%%%%%%%%%%%%%%%%%%%%%%%%%%%%%%%%%%%%%%%%%%%%%%%%%%%%%%%%%%%%%%%%%%%%%%%%%%%%%%%%%%

Flux compactifications constitute a promising direction to look for
string vacua with stabilised moduli \cite{Grana:2005jc,Douglas:2006es}.
However,  most of the supersymmetric solutions  found  so far feature a vacuum
with negative cosmological constant,
while Minkowski vacua with all the moduli stabilised in a perturbative
regime have remained elusive~\footnote{For a recent world-sheet approach for models with no K\"ahler
  moduli at strong coupling see \cite{Becker:2006ks}.  Also see
  \cite{Krefl:2006vu} for a non-perturbative possibility.}.

% in general, supersymmetric solutions feature a vacuum with negative
%cosmological constant which is clearly in contrast with present observations.
%Even if Minkowski solutions are allowed by supersymmetry, such vacua with all
%the moduli stabilised in a perturbative regime have remained
%elusive.\footnote{For a recent world-sheet approach for models with no K\"ahler
%  moduli at strong coupling see \cite{Becker:2006ks}.  Also see
%  \cite{Krefl:2006vu} for a non-perturbative possibility.}

%% Flux compactifications in string theory provide a rich landscape of vacua where all the moduli are stabilised \cite{Grana:2005jc,Douglas:2006es}.   
%% However, Minkowski vacua with all the moduli stabilised in a perturbative regime have remained elusive\footnote{For a recent world-sheet approach
%%  for models with no K\"ahler moduli at strong coupling see \cite{Becker:2006ks}.  Also see \cite{Krefl:2006vu} for a non-perturbative possibility.}. 
%% In this paper we outline a class of compactifications of both type IIA and type IIB string theory that lead to a vacuum where all the moduli are stabilised 
%% in a perturbative regime. 

On the other hand,
in the last years it  became clear that the original Calabi--Yau
compactifications (even with fluxes turned on) represent only a fraction
of all the possibilities which lead to supersymmetric theories
in lower dimensions. 
In particular, in Refs.~\cite{Grana:2004bg,Grana:2005sn}
it was shown that supersymmetric ground states in four dimensions are related
to internal manifolds with $SU(3)\times SU(3)$ structure which are called
\emph{twisted generalised Calabi--Yau} manifolds.

%% It is important to keep in
%% mind that these solutions only refer to the vacuum of the theory and for many
%% purposes it is interesting to consider more general backgrounds. In
%% particular, the whole picture of string dualities only seems to make sense if
%% general backgrounds with $SU(3) \times SU(3)$ structure are allowed.

Another important idea that allows to extend the class
of manifolds suitable for compactification is that of duality.
%
%It
%turns out that the most general solution are manifolds with $SU(3) \times
%SU(3)$ structure \cite{Grana:2004bg,Grana:2005sn,Grana:2005ny}. Such manifolds
%were not very much studied in the mathematics literature and before a complete
%theory for these manifolds -- which would allow us to perform the
%compactification from zero principles -- exists, the only avenue for exploring
%such compactifications is via dualities. 
For example, it is well established that type IIA/IIB string theories are
related by T-duality/mirror symmetry, in the absence of fluxes.  By insisting
  that this remains true
when fluxes are turned on, we should be able to find new
 configurations that are the  geometric
dual of  fluxes. In particular, it is expected
 that  the NS-NS fluxes play a major role,  as
it is well known that T-duality mixes the metric with the $B$-field:
 % and
%therefore one expects that the corresponding fluxes have some geometric
%dual. 
this was indeed confirmed in
Refs.~\cite{Gurrieri:2002wz,Kachru:2002sk,Gurrieri:2002iw} where it was shown
that the mirror/T --dual of the electric NS-NS fluxes are manifolds with
torsion. 
%% Since the original Calabi-Yau compactifications, the spectrum of possible compactification manifolds that are solutions of string theory has been 
%% growing. An important idea in extending the class of manifolds is that of dualities. Type IIA and IIB string theories are related by a T-duality which, at 
%% the four-dimensional level, manifests itself as mirror symmetry between the two four-dimensional theories that result from compactifications 
%% on manifolds that are T-dual to each other. We therefore expect that for each four-dimensional action obtained from a Calabi-Yau compactification with 
%% non-vanishing fluxes there should be a mirror type IIA compactification. It was shown that in the presence of electric NS fluxes the mirror 
%% manifolds required to obtain the correct four-dimensional theory are not Calabi-Yau but rather manifolds with torsion \cite{Gurrieri:2002wz,Kachru:2002sk}. 
%% 
These were half-flat manifolds, of which twisted-tori are a subclass. Compactifications on these type of manifolds was consequently studied in
\cite{Gurrieri:2002iw}-\cite{Palti:2006yz}
%\cite{Derendinger:2004jn,Villadoro:2005cu,House:2005yc,Camara:2005dc,Grana:2005ny,Acharya:2006ne,Curio:2000dw,Becker:2002sx,Curio:2003ur,Micu:2006ey,Gurrieri:2004dt,Micu:2004tz,deCarlos:2005kh,Manousselis:2005xa,Anguelova:2006qf,Gurrieri:2002iw,Behrndt:2005bv,Palti:2006yz}. 
It was realised that these 
type of manifolds induced superpotentials that had vacua where all the moduli were stabilised, all be it in an anti-deSitter vacuum.

The question of what type of manifolds are needed to recover the mirror in the presence of magnetic NS fluxes however remained a little more obscured. 
It was realised that these compactifications are not geometric in the usual sense of geometry Such non-geometric manifolds were studied in
\cite{Becker:2006ks,Kachru:2002sk}, \cite{Grana:2006is}--\cite{Persson:2006rd}
%\cite{Becker:2006ks,Grana:2006is,Benmachiche:2006df,Grana:2006hr,Kachru:2002sk,Hellerman:2002ax,Dabholkar:2002sy,Lowe:2003qy,Flournoy:2004vn,Kapustin:2004gv,Mathai:2004qc,Mathai:2004qq,Bouwknegt:2004ap,Hull:2004in,Hull:2005hk,Shelton:2005cf,Gray:2005ea,Dabholkar:2005ve,Lawrence:2006ma,Hull:2006tp,Hull:2006qs,Hull:2006va,Shelton:2006fd,Ellwood:2006my,Grange:2006es,Ellwood:2006ya,Persson:2006rd}
and the mirror to the magnetic NS fluxes were termed non-geometric fluxes. 
It was also conjectured in \cite{Grana:2005ny} that possible 
compactifications that would lead to the mirrors of magnetic NS fluxes are
compactifications on manifolds described by generalised geometry. This was
made more precise in \cite{Benmachiche:2006df,Grana:2006hr}.  

\smallskip

In this paper we study compactifications of type IIA and IIB string theories
on (non-geometric) orientifolds with $SU(3) \times SU(3)$ structure with the
purpose of finding four-dimensional supersymmetric Minkowski solutions.
%Manifolds with $SU(3) \times SU(3)$ structure  were first 
% For
%this we first need a way to derive the low-energy effective action for the
%corresponding compactifications. This was also recently addressed in
%ref.~\cite{Grana:2006hr}, but here we use a somehow different approach.
We argue that the formalism of generalised geometry allows us to treat
 non-geometric compactifications in a way very similar to geometric 
compactifications. Indeed by generalising the 
derivative operator to a covariant derivative for T-dualities we show that we are able to derive (a subclass of) the four-dimensional superpotentials of 
\cite{Shelton:2006fd,Aldazabal:2006up} from a compactification. 
Within this formalism we are able to derive the superpotential for an
arbitrary number of moduli. This is a crucial step toward finding Minkowski
vacua. We show that, under mild constraints on the numbers of moduli,
the superpotential we consider exhibits a rich spectrum of Minkowski vacua
with all the moduli stabilised and tadpole conditions satisfied.  

The paper outline is as follows. In section \ref{sec:gengeo} we introduce some concepts in generalised geometry and describe how we deal with the 
non-geometric nature of the compactifications. In section \ref{sec:comp} we derive the superpotentials and tadpole constraints for compactifications of 
type IIA and type IIB string theory on orientifolds with $SU(3) \times SU(3)$ structure.
In section \ref{sec:susymink} we study the vacuum structure of the superpotential for a number of simple cases and show that 
generically the vacuum spectrum can include Minkowski vacua with all the moduli stabilised at parametrically controlled values. 

%%%%%%%%%%%%%%%%%%%%%%%%%%%%%%%%%%%%%%%%%%%%%%%%%%%%%%%%%%%%%%%%%%%%%%%%%%%%%%%%%%%%%%%%%%%%%%%%%%%%%%%%%
\section{Some generalised geometry}
\label{sec:gengeo}
%%%%%%%%%%%%%%%%%%%%%%%%%%%%%%%%%%%%%%%%%%%%%%%%%%%%%%%%%%%%%%%%%%%%%%%%%%%%%%%%%%%%%%%%%%%%%%%%%%%%%%%%%%

In this section we give a brief overview of the parts of generalised geometry that are relevant for this work. We use the term `generalised geometry' 
to denote both generalised complex geometry and generalised almost complex geometry. Generalised geometry has been 
developed by mathematicians \cite{GCY,Hitchin:2005cv,Hitchin:2005in,CS,Gualtieri,Cavalcanti,Witt} and almost in parallel applied to Physics. By now there are a number of excellent reviews for physicists \cite{Grana:2005jc,Grana:2006is,Grana:2006hr,Grana:2006kf} which we follow for some of this section. 
In the first part of this section we introduce the objects and notations that we use in this paper. Following this we discuss an extension to the formalism in the form of a new derivative operator that can be thought of as gauging transformations of generalised complex structures. This kind of derivative has also been recently proposed in \cite{Grana:2006hr}. We end the section with a discussion of the some of the physics associated with generalised geometry.

%%%%%%%%%%%%%%%%%%%%%%%%%%%%%%%%%%%%%%%%%%%%%%%%%%%%%%%%%%%%%%%%%%%%%%%%%%%%%%%%%%%%%%%%%%%%%%%%%%%%%%%%%
\subsection{Generalised complex geometry}
\label{sec:gengeocom}
%%%%%%%%%%%%%%%%%%%%%%%%%%%%%%%%%%%%%%%%%%%%%%%%%%%%%%%%%%%%%%%%%%%%%%%%%%%%%%%%%%%%%%%%%%%%%%%%%%%%%%%%%%

Generalised complex geometry is the generalisation of complex geometry to $T\oplus T^*$, the tangent and cotangent bundles. 
An element of $T\oplus T^*$ is in general a sum of a vector $X$ and a one-form $\xi$. It is useful to write quantities in terms of matrices and vectors where the rows and columns denote whether the elements are in $T$ or in
 $T^*$. Then a generalised almost complex structure can be defined as a map $\J: T \oplus T^* \rightarrow T \oplus T^*$ that squares to $-\one_{2d}$, where $d$ is the real dimension of the manifold, and satisfies a Hermiticity condition $\J^T \I \J = \I$ where
\be
\I = \left(  \begin{array}{cc} 0  & \one_d \\ \one_d & 0 \end{array} \right) \;.
\ee 
An example of a generalised almost complex structure is one induced by an almost complex structure $I$ 
\be
\J_I = \left(  \begin{array}{cc} I  & 0 \\ 0 & -I^T \end{array} \right) \;. \label{gcscs}
\ee
Another example is one that is induced by an almost symplectic two-form $J$ 
\be
\J_J = \left(  \begin{array}{cc} 0  & -J^{-1} \\ J & 0 \end{array} \right) \;. \label{gcssi}
\ee
In general, a generalised almost complex structure will be some combination of the two. Given a generalised almost complex structure it is possible to generate a new one by a $B$-transformation defined as
\be
\J_B = \left(  \begin{array}{cc} 1 & 0 \\ B & 1 \end{array} \right) \J \left(  \begin{array}{cc} 1  & 0 \\ -B & 1 \end{array} \right) \;, \label{btransj}
\ee
where $B$ is a real two-form, $B \in \Lambda^2T^*$. Similarly we can also generate a new one through a $\beta$-transform
\be
\J_{\beta} = \left(  \begin{array}{cc} 1 & \beta \\ 0 & 1 \end{array} \right) \J \left(  \begin{array}{cc} 1  & -\beta \\ 0 & 1 \end{array} \right) \;,
\ee
where $\beta$ is a bivector, $\beta \in \Lambda^2T$. Given two generalised
almost complex structures, $\J_a$ and $\J_b$ that are compatible (ie they
commute) we also have a positive definite generalised metric 
\be
\G = - \J_a \J_b \;.
\ee
The two generalised almost complex structures reduce the structure group of the metric from $O(d,d)$ to $U\left(\frac{d}{2}\right)\times U\left(\frac{d}{2}\right)$. 

Consider a a generalised complex structure $\J$. Then since $\J^2 = -\one$ we can define projectors $\half \left( 1 \pm i\J\right)$ that split $T\oplus T^*$ into two subspaces $L$ and $\bar{L}$ that correspond to the $+i$ and $-i$ eigenvalues of $\J$. The subspace $L$ is maximal isotropic. Isotropic means that for 
any $v,w \in L$ we have 
\be
\left< v , w \right> = 0 \;,
\ee
where $\left<...,...\right>$ is the natural bracket on $T\oplus T^*$ defined as 
\be
\left<X+\xi,Y+\eta\right> = \half \left( \xi(Y) + \eta(X)\right) \;,
\ee
for general elements $X+\xi \in T \oplus T^*$. Maximal refers to the dimension of the subspace being the maximum value on the manifold, which is the dimension of the manifold. Then for each $\J$ there is a unique splitting into $L$ and $\bar{L}$ and conversely, a generalised complex structure is equivalent to a 
maximal isotropic subspace $L \subset \left( T \oplus T^* \right) \otimes \mathbb{C}$ such that $L \cap \bar{L} = \{0\} $. 
The type, $k$, of a maximal isotropic is the co-dimension of its projection onto $T$. Note that $B$-transforms do not change the type. In fact all the maximal isotropic subspaces are related by $B$-transforms. $\beta$-transforms can change the type by an even number.

%%%%%%%%%%%%%%%%%%%%%%%%%%%%%%%%%%%%%%%%%%%%%%%%%%%%%%%%%%%%%%%%%%%%%%%%%%%%%%%%%%%%%%%%%%%%%%%%%%%%%%%%%
\subsection{Pure spinors and $SU(3)\times SU(3)$ structure}
\label{sec:su3su3}
%%%%%%%%%%%%%%%%%%%%%%%%%%%%%%%%%%%%%%%%%%%%%%%%%%%%%%%%%%%%%%%%%%%%%%%%%%%%%%%%%%%%%%%%%%%%%%%%%%%%%%%%%%

The generalised almost complex structure defined in the previous section can also be expressed in terms of a pure spinor $\Phi$. This comes from the fact 
that the subspace $L$, of annihilators of a pure spinor $\Phi$
\be
\left( X + \xi \right) \Phi = 0 \;, \quad X + \xi \in L \; ,
\ee
is exactly a maximal isotropic subspace. The spinor can in turn be thought of as a formal sum of forms 
of even or odd degrees (corresponding to pure spinors with positive and negative chirality respectively). This is just the generalisation of generating a 
sum of forms from a spinor by acting on it with gamma matrices. In terms of the spinor $B$-transforms and $\beta$-transforms act as 
$\Phi \to e^B \Phi$ and $\Phi \to e^{\beta}\Phi$ respectively. Just as the existence of two generalised almost complex structures reduced the metric structure group to 
$U\left(\frac{d}{2}\right)\times U\left(\frac{d}{2}\right)$, the existence of two compatible (commuting) non-vanishing pure spinors reduces 
the structure group to $SU\left(\frac{d}{2}\right)\times
SU\left(\frac{d}{2}\right)$ where the extra reduction comes from the
non-vanishing constraint.

In this paper we are concerned with compactifications to four dimensions and so require a structure group $SU(3) \times SU(3)$.
Since each spinor also provides a supersymmetry parameter we see that compactifications on manifolds with $SU(3) \times SU(3)$ structure 
lead to $\N=2$ supersymmetry in four dimensions. 

It is possible to prove \cite{Gualtieri} that a pure spinor must always take the form
\be
\Phi = e^{B+iJ} \wedge \Omega_k \;,
\ee
where $B$ and $J$ are real two-forms and $\Omega$ is a complex form of degree $k$. $k$ is the type of $\Phi$ or of the generalised almost complex 
structure. Then a type 0 spinor is a $B$-transform of $e^{iJ}$ which in turn has non-zero norm if $J$ is non-degenerate and is closed if $dJ=0$. Hence 
it corresponds to a generalised complex structure that is derived from a symplectic structure as in (\ref{gcssi}). 
Similarly a type 3 spinor corresponds to a 
generalised almost complex structure derived from a complex structure as in (\ref{gcscs}). Then a general pure spinor is some combination of the two. 
We note here that in general the type of the spinor may vary throughout the manifold jumping by even numbers at particular loci on the manifold. It can 
also be seen that a $B$-transform preserves the type of the spinor while a $\beta$-transform in general need not.

%%%%%%%%%%%%%%%%%%%%%%%%%%%%%%%%%%%%%%%%%%%%%%%%%%%%%%%%%%%%%%%%%%%%%%%%%%%%%%%%%%%%%%%%%%%%%%%%%%%%%%%%%
\subsubsection*{An example: $\mathbb{C}_2$}
\label{sec:c2}
%%%%%%%%%%%%%%%%%%%%%%%%%%%%%%%%%%%%%%%%%%%%%%%%%%%%%%%%%%%%%%%%%%%%%%%%%%%%%%%%%%%%%%%%%%%%%%%%%%%%%%%%%%

In order to illustrate
  the notions defined above we consider the simple model of
$\mathbb{C}_2$ 

In this section we outline a very simple example of how the constructions described in the previous sections can be realised on the manifold $\mathbb{C}_2$.
Consider the standard complex structure on $\mathbb{C}_2$ defined by the spinor
\be
\Phi = dz_1 \w dz_2 \;.
\ee
In terms of the matrix notation the generalised complex structure reads
\be
\J = \left(  \begin{array}{cccc} 0 & 1 & 0  & 0 \\ -1 & 0 & 0 & 0 \\ 0 & 0 & 0 & 1 \\ 0 & 0 & -1 & 0 \end{array} \right) \;.
\ee
Now we may perform a $\beta$ transform. Since we have a full generalised complex structure rather than a generalised almost complex structure we should 
take the bivector to be holomorphic $\beta = z_1 \partial_{z_1} \w \partial_{z_2}$. Then performing a transform we find a new spinor
\be
\Phi_{\beta} = z_1 + dz_1 \w dz_2 \;. 
\ee
This is still a generalised complex structure, but now it is not simply induced by a complex structure. Rather it exhibits the jumping phenomenon between 
a generalised complex structure induced by a complex structure and one induced by a symplectic structure. To see this note that at the locus $z_1=0$ 
we recover the original type 2 generalised complex structure. Away from this locus we can write the spinor as
\be
\Phi_{\beta} = z_1e^{\frac{dz_1 \w dz_2}{z_1}} = z_1 e^{B+iJ}\;,
\ee
which is a ($B$-transform of a) generalised complex structure of type 0, induced by a symplectic structure on the manifold.

%%%%%%%%%%%%%%%%%%%%%%%%%%%%%%%%%%%%%%%%%%%%%%%%%%%%%%%%%%%%%%%%%%%%%%%%%%%%%%%%%%%%%%%%%%%%%%%%%%%%%%%%%
\subsection{Generalised geometry and non-geometry}
\label{sec:nongeo}
%%%%%%%%%%%%%%%%%%%%%%%%%%%%%%%%%%%%%%%%%%%%%%%%%%%%%%%%%%%%%%%%%%%%%%%%%%%%%%%%%%%%%%%%%%%%%%%%%%%%%%%%%%

In this section we outline the connection between generalised geometry as described in the previous sections and the physical view of non-geometry. The 
motivation
  is to reconcile the work in \cite{Kachru:2002sk,Hull:2004in} and \cite{Grange:2006es} with the discussion of the previous sections.

We approach non-geometry by considering a $T^2$ fibration over an $S^1$ given by the metric,
\be
ds^2 = dx^2 + dy^2 + dz^2 \;,
\ee
where $x$ is the co-ordinate along the circle and $y,z$ are the co-ordinates on the $T^2$. As we go around the $S^1$ we can identify the torus co-ordinates 
up to a $SL(2,\mathbb{Z})$ transformation which is the symmetry group of the torus. 
 Now we put $m$ units
 of $H$ flux through the manifold with a $B$-field given by 
\be
B \,=\, m\,x\, dy \w dz \;.
\ee
It was shown in \cite{Gurrieri:2002wz,Kachru:2002sk} that if we perform a T-duality along one of torus direction we reach a twisted-torus where now the manifold has 
torsion but there is no NS flux. The NS flux has been exchanged for metric fluxes. 
Performing another T-duality we reach the configuration
\ba
ds^2 &=& \frac{1}{1+m^2x^2}\left( dy^2 + dz^2 \right) + dx^2 \;, \nn \\ 
B &=& \frac{mx}{1+m^2x^2} dy \w dz \;.
\ea
We now see that if we go around the $S^1$ the $T^2$ metric is not periodic under an $SL(2,\mathbb{Z})$ transformation but rather under $O(2,2,\mathbb{Z})$. 
This new symmetry group corresponds to the geometric $SL(2,\mathbb{Z})$ plus T-dualities. Therefore this background can not be described geometrically in the 
usual sense. It is termed non-geometric or a T-fold. 

In \cite{Hull:2004in} a way of dealing with non-geometric manifolds was proposed. The basic idea is to double the number of dimensions so that 
$T^2 \ra T^4$, or more generally $T^n \ra T^{2n}$ for an $n$ torus. Then the geometric symmetry group becomes 
$SL(2 n,\mathbb{Z})$. The extra dimensions, or degrees 
of freedom, are then eliminated by a constraint which breaks the symmetry group down to $O(n,n,\mathbb{Z})$. 
Now this group is large enough to accommodate the T-dualities as well. 
The way this is realised in string theory is as follows. For a compactification on $T^n$, the internal momenta $P^i$, $i=1,..,n$, combine with the winding modes $w^i $ to form the momenta $P^i_L = P^i + w^i$ and $P^i_R = P^i - w^i$. Then the conjugate coordinates are written as 
$X^i_L = X^i + \tilde{X}^i$ and $X^i_R = X^i - \tilde{X}^i$, and $X^i$ are T-dual to $\tilde{X}^i$. The doubled formalism treats this as a compactification on $T^{2n}$ with coordinates $X$ and $\tilde{X}$, and the constraint that halves the degrees of freedom is imposing that $X_L$ and $X_R$ are 
left-moving and right-moving coordinates. Practically, this can be done by considering a projection of the doubled torus to 
a product torus $T^{2n} \ra T^n \oplus \tilde{T}^n$ such that the torus $T^n$ is identified with space-time and $X$ are the space-time coordinates. 
The CFT on the world sheet is invariant under which projection, or polarisation, we choose to make, i.e. we can 
rotate the fields $X$ and $\tilde{X}$ into each other. Then T-duality can be viewed as a rotation on the projectors. 
The failure to patch a manifold with geometric transition functions then corresponds to a failure to have a global polarisation.
 
The connection to generalised geometry is that under the space-time projection $T^{2n} \ra T^n$ the tangent bundle of the double torus 
is projected to the sum of the cotangent and tangent bundles of the space-time torus $T(T^{2n})\rightarrow \left(T \oplus T^*\right)(T^n)$. 
Then choosing the space-time $T^n$ within the $T^{2n}$ is equivalent to choosing a maximal isotropic subspace or a pure spinor, which under the 
projection of the tangent bundle leads to a pure spinor, or a maximal isotropic subspace, or a generalised almost complex structure in $T \oplus T^*$. 
Then T-duality, which is a rotation of the projectors, is given by a rotation of $\J$. More precisely 
the full symmetry group $O(n,n,\mathbb{Z})$ is simply the symmetry group of the generalised metric, elements of which can be split into three types 
$\left\{\beta, \varphi, B\right\}$ where $\varphi$ are elements of the geometric $SL(n,\mathbb{Z})$ subgroup. 
Then $B$ and $\beta$-transforms are non-geometric 
symmetries. The $B$-transforms are identified with shifts of the $B$-field that mix the metric and $B$-fields, 
and the $\beta$-transforms are identified with T-dualities.
Patching a manifold with T-dualities then corresponds to patching with $\beta$-transforms. This way we have a (generalised) geometric formulation 
of non-geometry.

There is another side to non-geometry that we have not mention so far. This is the issue of non-commutativity. It was shown in \cite{Kapustin:2003sg} 
that performing two T-dualities with flux leads to a non-commutative space. This was related to generalised geometry in 
\cite{Kapustin:2003sg,Grange:2006es} where it was shown that the non-commutativity 
is felt by open strings that end on separate patches which are patched by T-dualities. 
We note then that this non-commutativity only affects the 
open string sector and therefore does not alter the analysis in this paper.
It is worth mentioning that there are other non-geometric fluxes, denoted by $R$ in the literature \cite{Shelton:2005cf}, that arise from 
three T-dualities along directions with H-flux . These lead to non-associative manifolds which are non-geometric even locally.

%%%%%%%%%%%%%%%%%%%%%%%%%%%%%%%%%%%%%%%%%%%%%%%%%%%%%%%%%%%%%%%%%%%%%%%%%%%%%%%%%%%%%%%%%%%%%%%%%%%%%%%%%
\subsection{A supergravity analysis}
\label{sec:gautra}
%%%%%%%%%%%%%%%%%%%%%%%%%%%%%%%%%%%%%%%%%%%%%%%%%%%%%%%%%%%%%%%%%%%%%%%%%%%%%%%%%%%%%%%%%%%%%%%%%%%%%%%%%%

In this paper we perform a classical supergravity analysis of non-geometric compactifications. Given that non-geometric T-folds are inherently 
stringy in nature, a supergravity analysis requires some justification. In this section we present some reasoning in support of our approach and argue that 
our analysis may capture true vacua of the full string theory. We follow in part the discussions presented in \cite{Hull:2004in,Grana:2006hr}.

We begin this section with a brief discussion based on mirror symmetry which justifies the supergravity approach. This simply follows from the 
fact that the mirror duals of non-geometric compactifications are geometric. They can be as simple as a Calabi-Yau compactification with some H-flux. 
These compactifications are under control from a supergravity perspective and so we would expect that under the mirror identification 
their duals would also be valid\footnote{In this paper we require non-geometric fluxes to find Minkowski vacua on both the IIA and IIB sides of the mirror. 
However, for each case the non-geometric fluxes are dual to geometric fluxes. The IIA non-geometric fluxes are dual to magnetic NS flux, and the IIB 
non-geometric fluxes are dual to metric fluxes. Therefore we expect that the mirror symmetry reasoning presented above still holds.}.
The rest of this section is, in a sense, dedicated to suggesting 
 why the mirror symmetry works. That is to explaining why, by using generalised geometry,
 we can recover the IIA (non-geometric) mirrors of magnetic NS fluxes through a supergravity compactification. 

The first problem that a supergravity analysis of a T-fold faces is that the T-dualities that patch the T-fold mix momentum and winding modes. 
The ten-dimensional supergravity is derived by integrating out all the winding modes and 
keeping the momentum modes. On any particular patch this is valid, in the language of the doubled formalism we are choosing a polarisation. Globally, 
such a truncation is not valid since on different patches winding modes may become lighter than momentum modes following a T-duality. 
However, consider compactifying the ten-dimensional supergravity on a T-fold. 
To derive a four-dimensional effective theory we must now also integrate out the momentum modes, keeping 
only the zero mode (which has no winding equivalent). Therefore, given such a truncation of both the winding and momentum modes, any mixing 
should not affect the four-dimensional theory. In terms of this reasoning the energy scale of the low energy theory need not be any lower than 
in a geometric compactification since, if we are below the scale of a KK mode on a radius $R$, we are also below the scale of a winding mode on a 
radius $1/R$.

It is possible that it is the symmetry of the effective four-dimensional theory under T-dualities of the internal manifold 
that means the four-dimensional supergravity 
is the mirror to a supergravity derived from a geometric compactification. To see how this symmetry manifests itself in terms of the 
four-dimensional superpotential consider compactifying on an $SU(3) \times SU(3)$ structure manifold that is patched with $B$ and $\beta$-transforms. 
As outlined in section \ref{sec:nongeo}, this is a compactification on a T-fold. We have two pure spinors $\Phi^{\pm}$ on the manifold, and these would 
$B$ and $\beta$-transform between different patches. Let us turn off any RR-fluxes for simplicity. We decompose the ten-dimensional NS two-form 
$\hat{B}$ into a part that leads to the four-dimensional fields (axions) $B$ 
and a part responsible for background flux $B^{bg}$ so that the background 
flux $H^{bg}$ is given by $H^{bg}=dB^{bg}$ \footnote{This decomposition becomes more subtle in the presence of metric fluxes and is discussed in section 
\ref{sec:fluxax}.}. Then the four-dimensional superpotential was derived in \cite{Benmachiche:2006df} and reads
\be
W = \int_{M_6} \left< \left(d-H^{bg}\right) \tilde{\Phi}^+, \tilde{\Phi}^- \right> \;. \label{w4d1}
\ee 
where we absorbed the axions into the spinors $e^{-B}\Phi^{\pm} = \tilde{\Phi}^{\pm}$. 
The brackets $\left<,\right>$ denote the Mukai pairings and are 
defined in section \ref{sec:iiasu3su3}.
We can rewrite (\ref{w4d1}) in a suggestive way
\be
W = \int_{M_6} \left< d \left(e^{-B^{bg}}\tilde{\Phi}^+ \right), \left(e^{-B^{bg}} \tilde{\Phi}^- \right) \right> \;. \label{w4d2}
\ee
Written this way (\ref{w4d2}) is a $B$-transform of (\ref{w4d1}) with $-B^{bg}$ where we also transform the derivative
\be
d - H^{bg} \ra d - \left( H^{bg} + d\left( -B^{bg}\right) \right) = d \;.
\ee
Viewed in this way we can think of $B$-transforms as gauge transformations with a covariant derivative $\D_H = d-H$ that, under a $B$-transform transforms 
as $D_H \ra D_H - dB$. The superpotential 
\be
W = \int_{M_6} \left< \D_{H^{bg}} \tilde{\Phi}^+, \tilde{\Phi}^- \right> \;, \label{w4d3}
\ee
is then invariant under $B$-transforms and so can be integrated over a manifold patched with $B$-transforms to give the same four-dimensional theory.

Now we can also `gauge' $\beta$-transforms in the same way. The resulting covariant derivative is then 
\be
\label{tcd}
\D_{H,Q} = d - H - Q \;,
\ee
where we have introduced a new 'flux' which is a tensor that under a $\beta$-transform transforms as 
\be
Q \ra Q + d \w \beta \;.
\ee
Indeed the transformation of $Q$ corresponds to the change of the derivative operator under $\beta$-transforms as shown in \cite{Grange:2006es}, and 
can be thought of as encoding this change. It was also proposed in \cite{Grana:2006hr} by considering the possible fluxes that can be turned on.
We now have a covariant derivative of both $B$ and $\beta$-transforms
\be
\D_{H,Q}\Phi^{\pm} \ra e^{\beta+B}\left(\D_{H,Q}\Phi^{\pm} \right) \;,
\ee
so that the superpotential 
\be
W = \int_{M_6} \left< \D_{H^{bg},Q^{bg}} \tilde{\Phi}^+, \tilde{\Phi}^- \right> \;, \label{w4d4}
\ee
is invariant under both $B$ and $\beta$-transforms and so can be integrated over the internal manifold to give the four-dimensional theory.
The tensor $Q$ can be thought of as a flux in analogy with $H$ being the NS flux, and in the literature this would be identified with non-geometric flux. 
In general the operator $\D$ (we henceforth drop the indices) can send the degree of the form it acts on up one or three and if there is also $Q$-flux, 
down one. Finally we require the condition on the fluxes 
\be
i_Q \; H = 0 \;, \label{nilpo}
\ee
for the operator $\D$ to be nilpotent. 

We have seen that, using the covariant derivative $\D$, the ten-dimensional expression for the four-dimensional superpotential is invariant 
under $B$ and $\beta$-transforms\footnote{Note that also the K\"ahler potential is invariant since it is given by the Hitchin 
functional \cite{Hitchin:2005in,Grana:2005ny} 
$K = -i \mathrm{ln}\; \int_{M_6}{\left<\Phi^{\pm},\Phi^{\pm}\right> }$.}. To derive the effective four-dimensional superpotential in terms of the 
four-dimensional superfields we should decompose the spinors in terms of a basis of `low energy' forms. These would be the harmonic forms on a Calabi-Yau 
but more generally correspond to a truncation to a finite subset of forms
\cite{Grana:2005ny}. 
These forms would transform under B and $\beta$-transforms in the same way
as the spinors. However, as we have shown, such a transformation does not
change the expression for the superpotential which can still be written in
terms of the old basis forms.
Therefore the four-dimensional superpotential in terms of the 
superfields takes the same form on all the patches of the internal
T-fold.\footnote{Note that, as proposed in \cite{Grana:2006hr}, we could have
equally used a normal exterior derivative instead of $\D$ and considered the
basis forms to transform so that they become twisted 
$\omega \rightarrow (e^{\beta} \omega )$.}
%% \footnote{There is subtle point to note here. We have argued that we expected such 
%% a symmetry in the expression for the superpotential at the level of the KK zero mode level. However we have shown the ten-dimensional expression 
%% for the superpotential remains invariant even though we have not truncated to the zero modes, the pure spinors contain all the KK modes. 
%% The resolution to this may lie in the fact that the expression for the superpotential was derived from a ten-dimensional supergravity. This 
%% supergravity would not be valid globally at the ten-dimensional level because of the mixing of the winding and momentum modes. Instead we should 
%% think of a supergravity where all the higher momentum modes have also been truncated. We have performed the KK truncation already at the 
%% ten-dimensional level. This could explain why the resulting ten-dimensional expression has this symmetry.}. 

Having discussed the effects of the mixing of the winding and momentum modes let us briefly consider quantum effects. 
The mirror symmetry argument presented 
at the beginning of this section should still hold since mirror symmetry should be a quantum symmetry. In the case of non-perturbative 
effects such as instantons, gaugino condensation, light modes, the size of the effects  
 is determined by integrating the calibration form on the submanifolds they wrap. But these calibration forms are the pure spinors 
of the generalised geometry \cite{Gualtieri}. 
Therefore the size of these effects should be measured by the size of the four-dimensional superfields on each patch.
Note that, since these appear perturbatively in the superpotential, in vacua where they take large values, 
non-perturbative corrections to the superpotential will be highly suppressed.

Another possible correction source are $\alpha'$ corrections due to the
small curvature scales of the T-fold. These could come in the form of
corrections to the K\"ahler potential and corrections to the superpotential.
The corrections to the K\"ahler potential do not affect the supersymmetric
Minkowski vacua we are studying in this paper. Corrections to the
superpotential may be possible, but in the regime of large fields, which we
consider here, such corrections will still be subdominant.
%% but it would be difficult to see how to
%% reconcile these corrections with quantum mirror symmetry since the mirror
%% Calabi-Yau with H-flux superpotential should receive no such
%% corrections\footnote{For example, corrections to the cubic form of the
%% prepotential would still be dominated by the cubic term in the limit of
%% large field values.}.
Given this reasoning we proceed with the
compactifications neglecting any possible $\alpha'$ corrections.

%%  due to the small curvature scales in T-folds. We do not discuss these here 
%% since we do not have a clear understanding of how these transform under the mirror symmetry.
%% However we expect that these do not affect the supersymmetric 
%% Minkowski vacua we study since they are protected by non-renormalisation theorems \cite{Becker:2006ks}. 

To summarise, we have argued that it may be possible to treat the compactifications to Minkowski vacua we are studying in this paper as a four-dimensional supergravity. To remain in the perturbative regime we require the fields to take large values, for the dilaton this implies weak string coupling, 
for the geometric moduli the interpretation is a little less clear but the limit is required to suppress non-perturbative corrections. 
We also require to stay below the KK (or equally winding) scale. This is non-trivial to show in the presence of metric fluxes since the basis forms 
are no longer harmonic. However, we expect that there is such a low energy truncation since one exists in a Calabi-Yau mirror to a manifold 
with metric fluxes, see also \cite{Kashani-Poor:2006si,Grana:2005ny,Grana:2006hr} for constraints on these truncations. Therefore, assuming such a possible low energy truncation, 
we cautiously proceed with the supergravity analysis.

%%%%%%%%%%%%%%%%%%%%%%%%%%%%%%%%%%%%%%%%%%%%%%%%%%%%%%%%%%%%%%%%%%%%%%%%%%%%%%%%%%%%%%%%%%%%%%%%%%%%%%%%%
\section{The compactifications}
\label{sec:comp}
%%%%%%%%%%%%%%%%%%%%%%%%%%%%%%%%%%%%%%%%%%%%%%%%%%%%%%%%%%%%%%%%%%%%%%%%%%%%%%%%%%%%%%%%%%%%%%%%%%%%%%%%%%

In this section we derive the superpotential and tadpole constraints of the
 effective ${\cal N}=1$ theory resulting from compactifications of type II 
theories on orientifolds with $SU(3) \times SU(3)$ structure.
In both setups we consider an unwarped compactification given by the
 product Ansatz of the form $M_{10}\,=\,S_4 \times Y_6$. Here $S_4$
 is an unwarped
four dimensional space, while $Y_6$ is a six dimensional
compact manifold.
%
%
%\be
%\hat{g}_{MN}dX^{M}dX^{N} = g_{\mu\nu}dx^{\mu}dx^{\nu} + g_{mn}dy^m dy^n \;,
%\ee
%where $\hat{g}_{MN}$ denotes the ten-dimensional metric with co-ordinates $X^M$ which have an index range $M,N=0,...,9$. The $g_{\mu\nu}$ denotes the 
%external four-dimensional metric with co-ordinates $x^{\mu}$, $\mu,\nu=0,...,3$ and $g_{mn}$ is the internal metric with co-ordinates $y^m$, $m,n=1,...,6$.
%in the IIA case we denote the internal manifold $Y$ and in the IIB case we
% denote it $\tilde{Y}$.
 After deriving the two four-dimensional theories independently we give the mirror map under which they can be identified.

Throughout this paper we work in string units as in \cite{Aldazabal:2006up} where we take the basis forms $\alpha$,$\beta$,... to 
belong to an integer basis so that in units of $2\pi/\mu_{p-2}l^p = 1/l$, where $p$ is the degree of the form, $\mu_{p-2}$ is the Dp-brane unit of charge and $l=2\pi\sqrt{\alpha'}=1$ is the string length, the fluxes are integers. 

%%%%%%%%%%%%%%%%%%%%%%%%%%%%%%%%%%%%%%%%%%%%%%%%%%%%%%%%%%%%%%%%%%%%%%%%%%%%%%%%%%%%%%%%%%%%%%%%%%%%%%%%%
\subsection{Type IIA compactifications on $SU(3) \times SU(3)$ orientifolds}
\label{sec:iiasu3su3}
%%%%%%%%%%%%%%%%%%%%%%%%%%%%%%%%%%%%%%%%%%%%%%%%%%%%%%%%%%%%%%%%%%%%%%%%%%%%%%%%%%%%%%%%%%%%%%%%%%%%%%%%%%

The compactification of type IIA string theory on generalised orientifolds was studied in \cite{Benmachiche:2006df} and we follow their results in 
parts of this section. 
We begin by specifying the two compatible pure spinors $\Pi^{od}$ and $\Pi^{ev}$ 
\ba
\Pi^{od} &=& e^{-B} C\; \Omega \;, \\
\Pi^{ev} &=& e^{J_c} \;, \\
J_c & \equiv & -B + iJ \;,
\ea
Here $B$ denotes the (internal part of the) NS two-form that leads to the axions. It does not include the background contribution $B^{bg}$. 
$J$ is the K\"ahler form on the manifold, or more generally, 
for the non-K\"ahler cases we study, the almost symplectic two-form. We then see that $\Pi^{ev}$ is just a $B$-transform of the generalised almost complex 
structure induced by $J$. In general $\Pi^{od}$ can have type one, three or five and may jump between them. 
For the type three case $J$ and $\Omega$ are the $SU(3)$-structure forms that appear 
in standard $SU(3)$-structure compactifications (including Calabi-Yau compactifications). $C$ is the compensator field defined as 
\be
C \equiv e^{-\hat{\phi} - i\theta}e^{\frac12 (K^{cs}-K^K)} \;,\;\; e^{-K^{cs}}=i\Omega \wedge \bar{\Omega} \;,\;\; e^{-K^K} = \frac43 J \w J \w J \;, \label{Cdef}
\ee
where $\hat{\phi}$ is the ten-dimensional dilaton related to the four-dimensional dilaton $\phi$ through
\be
e^{-2\phi} = e^{-2\hat{\phi}} \; \frac43 \int_{Y}{J \w J \w J} \;.
\ee
The angle $\theta$ is fixed by the calibration condition for the orientifolds. In this paper we are interested in the compactifications that are 
mirror to type IIB compactifications with $O3/O7$ orientifolds in which case we choose the orientation of the $O6$ planes so that $\theta=0$.
The relevant ten-dimensional fields are completed by the RR fields which we define in formal sums
\be
C^{od} \equiv  C_1 + C_3 + C_5 + C_7 + C_9 \;. \label{iiacodd}
\ee
Note that only half the degrees of freedom in (\ref{iiacodd}) are physical. We also have the relevant field-strengths 
\be
F^{ev} = F_0 + F_2 + F_4 + F_6 \;,
\ee
where $F^{ev}=d_H C^{od}$. It is convenient to define the complex combination
\be
\Pi^{od}_c \equiv \tilde{C}_{(0)}^{od} + i\re{\Pi^{od}} \;,
\ee
where $\tilde{C}^{od}=e^{-B}C^{od}$, and the $(0)$ subscript denotes the component that is a four-dimensional scalar.
The four-dimensional superpotential resulting from this compactification was derived in \cite{Benmachiche:2006df} by reduction of the gravitino mass term. 
Here we use their expression but replace the twisted derivative with the full
covariant derivative $d_{H^{bg}} \ra \D$ as explained in the previous section.
With the definitions above the superpotential reads 
\be
W^{IIA} = \int_{Y}{ \left< \hat{F}^{ev}+ \D \Pi_c^{od} ,\Pi^{ev} \right> } \;. \label{wiia10d}
\ee
The brackets $\left< ... , ... \right>$ denote the Mukai pairing of forms defined as
\be
\left< \Psi^1 ,\Psi^2 \right> \equiv \Psi^1_{(0)} \w \Psi^2_{(6)} - \Psi^1_{(2)} \w \Psi^2_{(4)} + \Psi^1_{(4)} \w \Psi^2_{(2)} -
\Psi^1_{(6)} \w \Psi^2_{(0)} \;,
\ee
for a sums of even forms, and
\be
\left< \Psi^1 ,\Psi^2 \right> \equiv -\Psi^1_{(1)} \w \Psi^2_{(5)} + \Psi^1_{(3)} \w \Psi^2_{(3)} - \Psi^1_{(5)} \w \Psi^2_{(1)} \;,
\ee
for sums of odd forms. The subscripts denote the degree of the component of the forms. 
To derive the four-dimensional spectrum of fields we need to specify a finite set of basis forms on the manifold.
Following \cite{Grana:2005ny,Benmachiche:2006df} we consider a finite symplectic form basis
\ba
\omega_{\hat{A}} &=& \left( 1,\omega_{A}\right) \;,\;\; \tilde{\omega}^{\hat{A}} = \left( \tilde{\omega}^A, - \epsilon \right) \;, \\
\alpha_{\hat{K}} &=& \left( \alpha_0 , \alpha_{K} \right) \;,\;\; \beta^{\hat{K}} = \left( \beta_0, \beta^{K}  \right) \;.
\ea
The forms have non-vanishing Mukai pairings
\ba
\int_{Y} \left< \tilde{\omega}^{\hat{A}},\omega_{\hat{B}} \right> &=& \delta^{\hat{A}}_{\hat{B}} \;, \\
\int_{Y} \left< \alpha_{\hat{L}},\beta^{\hat{K}} \right> &=& \delta^{\hat{K}}_{\hat{L}} \;. \label{iiamuk}
\ea
In general the forms $\alpha_{\hat{K}}$ and $\beta^{\hat{K}}$ can have components that are one, three and five-forms.
However, since the expression for the superpotential does not depend on
which patch, or `gauge', we are on, we can always consider a local $SU(3)$
structure where $\alpha_{\hat{K}}$ and $\beta^{\hat{K}}$ are three-forms.
Under the orientifold projection the forms decompose into \cite{Benmachiche:2006df}
\ba
\left(1,\omega_a,\tilde{\omega}^{a},\epsilon \right) &\in & \Delta^{ev}_{+}  \;, \nn\\
\left( \omega_{\alpha}, \tilde{\omega}^{\alpha}\right) &\in & \Delta_{-}^{ev} \;, \nn \\
\left( \alpha_0,\alpha_{\lambda}, \beta^k \right) &\in & \Delta^{od}_{+} \;, \nn \\
\left( \alpha_k,\beta^{0},\beta^{\lambda} \right) &\in & \Delta^{od}_{-} \;. \label{iiabasis}
\ea
The sets $\Delta_{\pm}$ denote forms that are even or odd under the orientifold projection.
The index ranges are therefore $a=1,...,\mathrm{dim}\Delta^{2}_{+}$, $\alpha=1,...,\mathrm{dim}\Delta^{2}_{-}$. For the three-forms we consider the index 
ranges $k=1,...,\mathrm{dim}\Delta^{od}_{+}-1$ and $\lambda=\{ \o \}$. This is a choice we have in the definition of the basis forms \footnote{The more general case where we allow a non-trivial index range for $\lambda$ does not change anything in a fundamental way, rather we just need to turn on appropriate fluxes to recover the same superpotential. It makes the analysis more complicated however.}.
Under the orientifold splitting we have that 
\ba
\Pi^{ev} & \in & \Lambda^{ev}_{+} \;, \nn \\
\Pi^{od}_c & \in & \Lambda^{od}_{+} \;.
\ea
Therefore we decompose the forms into four-dimensional superfields as
\ba
\Pi^{ev} &=& 1 + iT^a \omega_a - \frac12 T^a T^b K_{abc} \tilde{\omega}^c - \frac{i}{6} T^a T^b T^c K_{abc} \epsilon \;, \nn  \\
\Pi^{od}_c &=& i S \alpha_0  + i U_k \beta^{k} \;, \label{iiapiev}
\ea
where we have introduced the analog of the intersection numbers
\be
K_{abc} \equiv \int_{Y} {\omega_a \w \omega_b \w \omega_c} \;.
\ee
We write the superfields in terms of real component fields as
\ba
T^a &=& t^a + ib^a \;, \\
U_k &=& u_k + i\nu_k \;, \\
S &=& s + i\sigma \;.
\ea
The imaginary components of the superfields are usually referred to as axions, a terminology that we keep, even though they all lose their axionic shift 
symmetries once fluxes are turned on. We refer to the fields $t^a$ as the K\"ahler moduli and to $u_k$ as the complex structure moduli, again keeping in 
mind that these manifolds need not be K\"ahler or complex. Finally the field $s$ is the related to the dilaton through the definition of the compensator field $C$.

The fluxes transform under the orientifold action such that 
\ba
F^{ev} &\in & \Lambda^{ev}_{+}\;, \nn \\
H &\in &  \Lambda^{3}_{-} \;.
\ea
Turning on background fluxes raises a delicate issue that is discussed in more detail in section \ref{sec:fluxax} where the distinction between a 
background flux and an axion vacuum expectation value is made more rigorous. For now we decompose the field-strengths $H$ and $F^{ev}$ as 
\ba
F^{ev} &=& F^{bg} + dC^{od} \;, \\
H &=& H^{bg} + dB \;,
\ea
and we set 
\ba
F^{bg} &=& - m^{0} + p^a \omega_a - q_a \tilde{\omega}^a - e_0 \epsilon \;, \\
H^{bg} &=& - h_0 \beta^0 +  h^k \alpha_k \;.
\ea
We take the background flux parameters to be integer quantised (ignoring any possible subtleties with half-integer values induced by orientifolds). 

To calculate the superpotential we should also specify the differential properties of the basis forms. In general they need not be harmonic, i.e. closed and co-closed. Indeed if there is torsion on the manifold they will not be. The torsion is usually referred to as metric fluxes, an 
appropriate term when it comes to looking at mirror symmetry where these fluxes are dual to electric NS flux. On our manifold we generally also 
have non-geometric $Q$-flux, this appears like metric fluxes but corresponds to the degree of the form decreasing by one. They are mirror dual to 
magnetic NS flux \cite{Grana:2005ny,Benmachiche:2006df,Grana:2006hr}. 
For the differential relations of the basis forms we take \footnote{We have included only the background contribution, $H^{bg}$, 
of the NS flux to the operator $\D$. This is consistent with our approach in section \ref{sec:fluxax}.} 
\ba
\D\alpha_{0} &=& - m^a \omega_a - e_{a} \tilde{\omega}^a - h_0 \epsilon \;, \nn \\
\D\beta^{k} &=& -e_a^{\;\;k} \tilde{\omega}^a - h^k \epsilon \;, \nn \\ 
\D\omega_a &=& - e_{a} \beta^0 + e_a^{\;\;k} \alpha_k \;, \nn \\
\D\tilde{\omega}^a &=& m^a \beta^{0} \;. \label{diffiia}
\ea
All other forms are closed. The fluxes $e_{a}$ and $e_a^{\;\;k}$ are the metric fluxes and $m^a$ are the non-geometric fluxes. These relation are not the 
most general relations, we have not turned on all the possible fluxes. Rather we have turned on the minimum amount of fluxes 
needed to find the Minkowski vacua. We note that we are in a sense allowing ourselves too much freedom. The metric, and non-geometric, 
fluxes will be fixed by the particular manifold we are choosing to compactify on. However, 
all our fluxes, metric and non-geometric, 
 have mirrors that are simply NS fluxes \cite{Gurrieri:2002wz,Benmachiche:2006df,Louis:2006kb,Grana:2006hr} 
and so should in principle be tunable even though we may not have an explicit manifold for every choice.

Substituting all this into the superpotential (\ref{wiia10d}), and integrating over the internal manifold, we find
\begin{eqnarray}
W^{IIA} &=& \frac{i}{6} m^{0} \,K_{a b c}\; T^{a} T^{b}  T^{c} + \frac12 K_{a b c} \;   p^{c}T^{a} T^{b}
-i q_{a} \,T^{a}  +  e_0 \nonumber\\ 
&-& \; \frac{i}{2} S K_{a b c}\,  m^{a} T^{b}  T^{c} + S e_{a} T^a + e_a^{\;\;k} U_k T^{a} + i h_0 S + i h^{k}  U_k \;.\label{wiiahf}
\end{eqnarray}
This type of superpotential was already proposed in \cite{Shelton:2006fd,Aldazabal:2006up} through mirror symmetry. 

%%%%%%%%%%%%%%%%%%%%%%%%%%%%%%%%%%%%%%%%%%%%%%%%%%%%%%%%%%%%%%%%%%%%%%
\subsubsection{IIA tadpoles and Bianchi identities}
%%%%%%%%%%%%%%%%%%%%%%%%%%%%%%%%%%%%%%%%%%%%%%%%%%%%%%%%%%%%%%%%%%%%%%

In this section we write the constraints that arise from tadpole constraints/Bianchi identities. 
In the RR sector we have the tadpole constraint/Bianchi identity \cite{Grana:2006kf}
\be
d_H F^{ev} = \delta_{\mathrm{source}} \;.
\ee
Again, we generalise this to 
\be
\D F^{ev} = \delta_{\mathrm{source}} \label{iiatad} \;.
\ee
The localised RR charged sources are denoted by $\delta_{\mathrm{source}}$ and can be wrapped $D6$ branes or $O6$ orientifolds. It is important to notice 
that the full field-strengths, rather than just the background ones, appear in the relations. Hence, in general, there is a contribution 
from the vacuum expectation values of the axions. This issue is discussed in more detail in section \ref{sec:fluxax} and for now we only include the 
background flux $H^{bg}$ in the expressions. Putting the expressions for the field-strengths into the 
Bianchi identity we find
\be
\left( - m^0 h_0 - p^a e_{a} - m^a q_a \right) \beta^0 - \left( m^0 h^k + p^a e_{a}^{\;\;k} \right) \alpha_k = \delta_{\mathrm{source}} \label{iiatad1}\;. 
\ee
Following \cite{Acharya:2006ne,Grana:2006kf} we proceed to 'smear' the localised sources so that (\ref{iiatad1}) can be solved for each component.
This involves replacing the localised sources with integer charges multiplying an appropriate form. 
The sign of the resolved forms is fixed by an appropriate calibration condition as in \cite{Grana:2006kf}. 
Where for a resolved form $\gamma$ we require \footnote{There is a minus sign with respect to the conventions of \cite{Grana:2006kf} due to the different definitions of the Mukai pairings.}
\be
\left<\im{\left(\Pi^{od}_c\right)},\gamma\right> \sim +\epsilon \;.
\ee
With the sign fixed in this way we reach the constraints
\ba
- m^0 h_0 - p^a e_{a} - m^a q_a &=& Q_0 \equiv 2N^{D6}_{0} - 4 N^{O6}_{0} \;, \\
- m^0 h^k - p^a e_{a}^{\;\;k} &=& Q^k \equiv  2 N^{k,\;D6} - 4 N^{k,\;O6} \;, \label{iiatad2}
\ea
where $N^{k,\;06/D6}$ denotes the number of $O6/D6$ planes wrapped on the submanifold $\beta^k$. 
The Bianchi identity for the NS flux reads $d H = 0$. Note that using the
condition for  $\D$ to be nilpotent, (\ref{nilpo}), the Bianchi identity can
be written as $\D H =0$ which is identically satisfied for our choice of
fluxes. 
%% and is satisfied identically for our choice of fluxes. It is interesting that this Bianchi identity 
%% implies the condition (\ref{nilpo}) which was required for $\D$ to be nilpotent.
Finally, the differential relations for the basis forms (\ref{diffiia}) impose a self consistency constraint, obtained by taking two 
derivatives
\be
m^a e_{a}^{\;\;k} = 0 \;. \label{iiatadme}
\ee

\noindent
Notice that we do not consider any non-Abelian gaugings to be generated by
  ((non)-geometric) fluxes and therefore the number of constraints
which we have for the parameters are less than in the twisted tori case
\footnote{
We thank Pablo Camara for pointing us this fact.}.

%%%%%%%%%%%%%%%%%%%%%%%%%%%%%%%%%%%%%%%%%%%%%%%%%%%%%%%%%%%%%%%%%%%%%%%%%%%%%%%%%%%%%%%%%%%%%%%%%%%%%%%%%
\subsection{Type IIB compactifications on $SU(3) \times SU(3)$ orientifolds}
\label{sec:iibsu3su3}
%%%%%%%%%%%%%%%%%%%%%%%%%%%%%%%%%%%%%%%%%%%%%%%%%%%%%%%%%%%%%%%%%%%%%%%%%%%%%%%%%%%%%%%%%%%%%%%%%%%%%%%%%%

These compactifications follow in a very similar way to the IIA compactifications discussed in section \ref{sec:iiasu3su3} and 
so we just outline the important steps. The orientifolds we consider are of the $O3/O7$ type. The compactification 
was also studied in \cite{Benmachiche:2006df}.

The pure spinors are 
\ba
\Phi^{od} &\equiv& e^{-B}\Omega \;,\\
\Phi^{ev} &\equiv& e^{-\hat{\phi}} e^{J_c} \;.
\ea
The RR form fields are all even and we define the formal sums
\ba
C^{ev} &\equiv& C_0 + C_2 + C_4 + C_6 + C_8 \;, \nn \\
F^{od} &\equiv& F_1 + F_3 + F_5 \;.
\ea
We form the complex superfield combination
\be
\Phi^{ev}_c \equiv C_{(0)}^{ev} + i\re{\Phi^{ev}} \;,
\ee
where $C_{(0)}^{ev}$ denotes the component of $C^{ev}$ that is a four-dimensional scalar.
With this the superpotential reads \cite{Benmachiche:2006df}
\be
W^{IIB} = \int_{\tilde{Y}}{\left< F_3 + \D \Phi^{ev}_c , \Phi^{odd}  \right>} \;, \label{wiibgen}
\ee
where $\tilde{Y}$ denotes the internal manifold, and 
 we have replaced $d_H \rightarrow \D$. In order to compute the four-dimensional spectrum we restrict to 
a finite symplectic basis of forms as in (\ref{iiabasis}). Under the orientifold projection these split as
\ba
\left(1,\omega_{\alpha},\tilde{\omega}^{a} \right) &\in & \Delta^{ev}_{+}  \;, \\
\left( \omega_{a}, \tilde{\omega}^{\alpha} ,\epsilon \right) &\in & \Delta_{-}^{ev} \;, \\
\left( \alpha_0, \alpha_{k}, \beta_0, \beta^{k} \right) &\in & \Delta^{od}_{-} \;, \\
\left( \alpha_{\lambda},\beta^{\lambda} \right) &\in & \Delta^{od}_{+} \;.
\ea
As in the IIA case we truncate the spectrum so that the index $\alpha = \{ \o \}$. Under the orientifold action we have that 
\ba
\Phi^{ev}_c & \in & \Lambda^{ev}_{+} \;, \nn \\
\Phi^{od} & \in & \Lambda^{od}_{-} \;,
\ea
and so we decompose the spinors into superfields as
\ba
\Phi_c^{ev} &=& i \tau - i T_a \tilde{\omega}^a \;, \nn \\
\Phi^{odd} &=& \alpha_{0} + i U^k \alpha_k  + \frac12 U^k U^l \kappa_{klm} \beta^m - \frac{i}{6} U^k U^l U^m \kappa_{klm} \beta^0 \;. \label{pureiib}
\ea
Here the coefficients $\kappa_{klm}$ feature in the prepotential for the complex structure moduli. 
We write the superfields in terms of their real component fields as
\ba
T_a &=& t_a + i b_a \;, \nn \\
U^k &=& u^k + i\nu^k \;, \nn \\
\tau &=& e^{-\hat{\phi}} + i\sigma \;.
\ea
We now turn to specifying the fluxes. The background fluxes we consider are 
\ba
H^{bg} &=& m^k \alpha_k - e_k \beta^k + h_0 \beta^0 \;, \nn \\ 
F_3^{bg} &=& - m^0 \alpha_0 + p^k \alpha_k + q_k \beta^k - e_0 \beta^0 \;.
\ea
The metric and non-geometric fluxes are specified in the differential relations for the forms which we take to be 
\ba
\D \alpha_0 &=& h^a \omega_a + h_0 \epsilon\;, \nn \\
\D \alpha_k &=& - e_{k}^{\;\;a} \omega_a - e_k \epsilon \;, \nn \\
\D \tilde{\omega}^a &=& - e^{\;\;a}_{k}\beta^k + h^a \beta^0 \;.
\ea
With these conventions the superpotential (\ref{wiibgen}) evaluates to 
\ba
W^{IIB} &=& \frac{i}{6} m^{0} \,\kappa_{klm}\; U^{k} U^{l}  U^{m} + \frac12 \kappa_{klm} \; p^{k}U^{l} U^{m}
- iq_{k} \,U^{k}  +  e_0 \nonumber\\ 
&-& \; \frac{i}{2} \tau \kappa_{klm}\;  m^{k} U^{l}  U^{m} + T_a e_{k}^{\;\;a} U^{k} + ih^a T_a + \tau e_k U^k + ih_0 \tau \label{wiibhf} \;.
\ea

%%%%%%%%%%%%%%%%%%%%%%%%%%%%%%%%%%%%%%%%%%%%%%%%%%%%%%%%%%%%%%%%%%%%%%
\subsubsection{IIB tadpoles and Bianchi identities}
%%%%%%%%%%%%%%%%%%%%%%%%%%%%%%%%%%%%%%%%%%%%%%%%%%%%%%%%%%%%%%%%%%%%%%

The Bianchi identities or tadpole conditions for the IIB case read \cite{Grana:2006kf}
\be
\D F^{od} = \delta_{source} \;.
\ee
Putting in the expressions for the field-strengths and regularising the sources we find 
\ba
-m^0 h_0 - m^k q_k - e_k p^k  &=& Q_0 \equiv 2 N^{D3} - \frac12 N^{O3} \;, \nn \\
-p^k e_k^{\;\;a} - m^0 h^a &=& Q^a \equiv 2 N^{a,\;D7} - 8 N^{a,\;O7}  \;.
\ea
We also have the constraint arising from the Bianchi identity for the NS field $\D H = 0$
\be
m^k e_k^{\;\;a} = 0 \;.
\ee

%%%%%%%%%%%%%%%%%%%%%%%%%%%%%%%%%%%%%%%%%%%%%%%%%%%%%%%%%%%%%%%%%%%%%%%%%%%%%%%%%%%%%%%%%%%%%%%%%%%%%%%%%
\subsection{The mirror map}
\label{sec:mirmap}
%%%%%%%%%%%%%%%%%%%%%%%%%%%%%%%%%%%%%%%%%%%%%%%%%%%%%%%%%%%%%%%%%%%%%%%%%%%%%%%%%%%%%%%%%%%%%%%%%%%%%%%%%%

The two four-dimensional theories derived in sections \ref{sec:iiasu3su3} and \ref{sec:iibsu3su3} were constructed as a mirror pair. 
It is clear that the superpotentials (\ref{wiiahf}) and (\ref{wiibhf}), and pure spinors (\ref{iiapiev}) and (\ref{pureiib}), 
match under the identification of the superfields
\be
\tau \leftrightarrow S \;,\;\; U^k \leftrightarrow T^a \;,\;\; U_k \leftrightarrow T_a \;,
\ee
and the basis forms
\be
1 \leftrightarrow \alpha_0 \;,\;\; \omega_a \leftrightarrow \alpha_k \;,\;\; \tilde{\omega}^a \leftrightarrow -\beta^k  
\;,\;\; \epsilon \leftrightarrow \beta^0   \;,
\ee
The mirror fluxes on each side are denoted by the same symbols. 
Note that on the IIB side $\tau$ is the ten-dimensional dilaton and on the IIA side $S$ is the 
compensator field defined in (\ref{Cdef}). The actual equivalence is at the four-dimensional dilaton level. The quantity we should keep large is the 
inverse string coupling which is the ten-dimensional dilaton
\be
g_s^{-1} = e^{-\hat{\phi}} \gg 1 \;.
\ee 

%%%%%%%%%%%%%%%%%%%%%%%%%%%%%%%%%%%%%%%%%%%%%%%%%%%%%%%%%%%%%%%%%%%%%%%%%%%%%%%%%%%%%%%%%%%%%%%%%%%%%%%%%
\subsubsection{Exact flux and axions}
\label{sec:fluxax}
%%%%%%%%%%%%%%%%%%%%%%%%%%%%%%%%%%%%%%%%%%%%%%%%%%%%%%%%%%%%%%%%%%%%%%%%%%%%%%%%%%%%%%%%%%%%%%%%%%%%%%%%%%

In section \ref{sec:iiasu3su3} we faced the problem of whether to include the
contribution from the axions, $b^a$, to the tadpole equations. Recall that due
to the non-closure of the forms in which we expand, a non-trivial vacuum
expectation value for the $B$-field on the internal manifold can lead to a
change in the $H$-flux. The tadpole conditions as stated in (\ref{iiatad2})
only include a contribution from a 'background' flux $H^{bg}$. From a
ten-dimensional point of view however, the Bianchi identities are not
sensitive to the splitting between the background flux and the vacuum
expectation values for the axions. This motivates the argument that the axions
vevs should feature in the tadpole conditions. 
With the axion contributions the tadpole conditions (\ref{iiatad2}) read 
\ba
\left( - m^0 e_{a} <b^a> \right) - m^0 h_0 - p^a e_{a} - m^a q_a &=& Q_0  \;, \nn \\
\left( - m^0 e_a^{\;\;k} <b^a> \right) - m^0 h^k - p^a e_{a}^{\;\;k} &=& Q^k  \;, \label{iiatad5}
\ea
where $<b^a>$ denoted the vacuum expectation value for the scalar field $b^a$.
Therefore, in a vacuum with non-vanishing combinations $e_{a} b^a$ or $ e_a^{\;\;k} b^a$, 
if we solve the conditions without the axions (\ref{iiatad2}) then we do not solve the 
ten-dimensional Bianchi identities.

Even if the argument above for taking into account the axion vevs into the
tadpole conditions sounds convincing, there are also other things one should
consider. One of the main problems to be addressed is the fact that on the
mirror IIB side this does not seem to have any obvious analog. This problem
arises even in the simplest cases of half-flat manifolds as in
\cite{Gurrieri:2002wz}. A possible resolution could be that we should only
identify the two theories in a particular vacuum, and this would be one where
the axion combinations $e_{a} b^a$ and $ e_a^{\;\;k} b^a$ vanish. As the IIB
mirrors to the NS axions are the real parts of the complex structure moduli,
it may be that (see footnote 23 in \cite{Grana:2006kf}) non trivial
values for them break isometries of the manifold that are required to perform 
the mirror symmetry. Another possible way out is if we consider the NS flux on
the IIB side to be sourced by NS5-branes wrapped on cycles dual to the ones
with flux. The calibration condition for the branes is then the mirror
condition to requiring that the axions on the IIA side vanish. 

%% On the other hand if we include the axion contributions to the tadpoles we face new problems. 
%% There is no analogous contribution to the tadpoles on the type IIB side and so the two theories no longer look 
%% mirror symmetric. This problem arises in even the simplest cases as in \cite{Gurrieri:2002wz}. Further, the tadpole conditions 
%% (\ref{iiatad5}) do not respect the integer axion shift symmetries of the superpotential. By these we mean that the superpotential 
%% (\ref{wiiahf}) has a symmetry where we transform $b^a \ra b^a + r^a$, where $r^a$ are some integers, and also transform the fluxes appropriately, 
%% see for example (\ref{axshex}). 

%% It is unclear how to resolve these issues. A possible resolution could be that we should only identify the two theories in a particular vacuum, 
%% and this would be one where the axion combinations $e_{a} b^a$ and $ e_a^{\;\;k} b^a$ vanish. The IIB mirrors to the NS axions are the real parts of the 
%% complex structure moduli. Then it may be that (see footnote 23 in \cite{Grana:2006kf}) non trivial values for them break isometries of the manifold that are 
%% required to perform the mirror symmetry. Another possible condition is, if we consider the NS flux on the IIB side to be sourced by 
%% NS5-branes wrapped on cycles dual to the ones with flux, then the calibration condition for the branes is mirror to requiring that the axions 
%% on the IIA side vanish. It may be that we should implement conditions such as these in the mirror identification. 
  
%Since none of the above resolutions is satisfactory 
In this work, we adopt  the following
a pragmatic approach that is correct in both cases and which is at most a  
superfluous constraint on the fluxes. We look for vacua where the combinations
$e_{a} b^a$ and $ e_a^{\;\;k} b^a$ vanish. We implement this practically 
by assuming that these combinations vanish and using the tadpole conditions
(\ref{iiatad2}) to solve for the fields. Then choose the fluxes so that  
in the vacuum this condition is satisfied thereby justifying our assumption. 
This procedure  is clarified in the next section, where we present explicit
solutions.  

%%%%%%%%%%%%%%%%%%%%%%%%%%%%%%%%%%%%%%%%%%%%%%%%%%%%%%%%%%%%%%%%%%%%%%%%%%%%%%%%%%%%%%%%%%%%%%%%%%%%%%%%%
\section{Supersymmetric Minkowski vacua}
\label{sec:susymink}
%%%%%%%%%%%%%%%%%%%%%%%%%%%%%%%%%%%%%%%%%%%%%%%%%%%%%%%%%%%%%%%%%%%%%%%%%%%%%%%%%%%%%%%%%%%%%%%%%%%%%%%%%%

In this section we analyse the vacuum structure of the superpotentials (\ref{wiiahf}) and (\ref{wiibhf}). 
We search for supersymmetric Minkowski vacua that are 
solutions to the equations
\be
\partial_{T^a} W = \partial_{U_k} W = \partial_{S} W  = W = 0 \;. \label{susyminkeq}
\ee
We begin this section with some no-go theorems regarding the existence of 
 physical Minkowski vacua in our set-up, placing constraints on the type of 
 manifold we should compactify on. In section \ref{sec:susyminkvac} we focus 
 on cases with a reduced number
 of fields, and study explicit realisations of Minkowski 
 vacua. Then, in section \ref{sec:genvac},  we provide
some arguments to tackle  the
 general case. Throughout this section we work in type IIA notation, with the
superpotential given in formula (\ref{wiiahf}). 

\smallskip

We note that in \cite{Camara:2005dc,Aldazabal:2006up,Shelton:2006fd} searches for Minkowski vacua were conducted and examples of supersymmetric 
Minkowski vacua were found but with not all of the moduli fixed. Some arguments regarding
constraints on fixing all the moduli, similar to the no-go theorems we are going to discuss,
 were consequently presented.

%%%%%%%%%%%%%%%%%%%%%%%%%%%%%%%%%%%%%%%%%%%%%%%%%%%%%%%%%%%%%%%%%%%%%%%%%%%%%%%%%%%%%%%%%%%%%%%%%%%%%%%%%
\subsection{Some no-go theorems}
\label{sec:nogo}
%%%%%%%%%%%%%%%%%%%%%%%%%%%%%%%%%%%%%%%%%%%%%%%%%%%%%%%%%%%%%%%%%%%%%%%%%%%%%%%%%%%%%%%%%%%%%%%%%%%%%%%%%%

In this section we note two conditions that the compactification manifold must satisfy for Minkowski vacua with all the moduli stabilised to exist.  
We work in a completely perturbative regime and all our statements are with respect to the superpotential (\ref{wiiahf}). More fluxes, non-perturbative or 
higher order effects would change the superpotential, and mean that our statements are not applicable.
In section \ref{sec:nominknogeo} we show that for the IIA superpotentials we are considering, 
in the absence of non-geometric fluxes there are no Minkowski vacua 
at finite volume with all the moduli stabilised
\footnote{The
 IIB case is much simpler, since  the dependence on Kahler 
moduli only comes from non-geometric fluxes.}.  
 In section \ref{sec:nominkkcs} we show that in IIA the number of K\"ahler moduli must be larger than the number of complex structure moduli in order 
for all the moduli to be stabilised, with the mirror statement also holding in IIB.

%%%%%%%%%%%%%%%%%%%%%%%%%%%%%%%%%%%%%%%%%%%%%%%%%%%%%%%%%%%%%%%%%%%%%%%%%%%%%%%%%%%%%%%%%%%%%%%%%%%%%%%%%
\subsubsection{No SUSY Minkowski with all moduli stabilised for IIA/O6 without non-geometric fluxes}
\label{sec:nominknogeo}
%%%%%%%%%%%%%%%%%%%%%%%%%%%%%%%%%%%%%%%%%%%%%%%%%%%%%%%%%%%%%%%%%%%%%%%%%%%%%%%%%%%%%%%%%%%%%%%%%%%%%%%%%%

We set out to show that without non-geometric fluxes it is impossible to stabilise all the moduli in a supersymmetric Minkowski vacuum.
The general superpotential we consider is given by (\ref{wiiahf}). In order 
to end with a Minkowski vacuum, we must impose (\ref{susyminkeq}).
Let us consider the following combination, that must vanish at the minimum we are interested in 
\be
\im{\left( t^a \partial_{T^a} W + u^k \partial_{U^k} W + s\partial_{S} W - W \right)} = 0\,. \label{imw0}
\ee
A straightforward computation using the superpotential \eqref{wiiahf} shows that the condition (\ref{imw0}) can be rewritten as
\be
\frac{m^0}{3} K_{abc} t^a t^b t^c = s K_{abc} m^a t^b t^c  \;.\label{conimpart}
\ee
If we consider vanishing non-geometric fluxes, $m^a=0$ for all $a$, the right hand side of (\ref{conimpart}) 
 vanishes. 
The left hand side is proportional to the volume of the compactification manifold. This implies that the only way to 
satisfy this relation is to choose $m^0=0$. This leaves us with a superpotential that is at most quadratic in the fields. 

To proceed with the argument, it is convenient to combine the superfields into one set of fields $\tilde{T}^{\Sigma} \equiv \left( T^{a},S,U^{k} \right)$, 
with index $\Sigma,\Lambda=1,...,\mathrm{dim}\Delta^{2}_{+} + \mathrm{dim}\Delta^{od}_{+}$. 
We decompose the fields into real and imaginary part as $\tilde{T}^{\Sigma}=\tilde{t}^{\Sigma}+i \tilde{\tau}^{\Sigma}$.
The superpotential (\ref{wiiahf}) can consequently be rewritten as
\be
W \,=\, b_{\Sigma\Lambda}\tilde{T}^{\Sigma}\tilde{T}^{\Lambda} + i c_{\Sigma} \tilde{T}^{\Sigma} - e_0 \;,
\ee
where $b_{\Sigma\Lambda}$ is a square, symmetric, real matrix and $c_{\Sigma}$ a real vector, both depending only on the flux parameters.
The condition for a supersymmetric vacuum in Minkowski space,
$\partial_{\tilde{T}^{\Sigma}} W = 0$ decomposed into real and imaginary parts
reads 
\begin{eqnarray}
  b_{\Sigma\Lambda}\tilde{t}^{\Lambda} & = & 0 \label{syssimca} \;, \\
  2 b_{\Sigma\Lambda}\tilde{\tau}^{\Lambda} + c_{\Sigma} & = &0
  \label{syssimcaim} \;. 
\end{eqnarray}
Contracting \eqref{syssimcaim} with $\tilde t^\Sigma$ and using
\eqref{syssimca} we obtain
\be
c_{\Sigma}\tilde{t}^{\Sigma}\,=\,0 \label{syssimcaim2}\,.
\ee
Note that these equations always have a flat direction as rescaling the fields
$\tilde t^\Sigma$ by some number leaves them unchanged. In principle this can
be resolved by the additional constraint $W=0$ which has to be satisfied in a
Minkowski vacuum. However it is easy to see that the above equations conspire
to eliminate any dependence of $W$ on $\tilde t$ at the critical point leaving
us with the flat direction. Finally, as this direction is a proper geometric
modulus it has no chance of being the QCD axion and therefore leaving it
unfixed is undesirable.  
%% Now, let us consider the real part of the condition $\partial_{\tilde{T}^{\Sigma}} W = 0$ which reads
%% \be
%% b_{\Sigma\Lambda}\tilde{t}^{\Lambda}\,=\,0 \; \label{syssimca} \;.
%% \ee
%% The imaginary part of $\partial_{\tilde{T}^{\Sigma}} W = 0$ is
%% \be
%% 2 b_{\Sigma\Lambda}\tilde{\tau}^{\Lambda} +
%% c_{\Sigma}\,=\,0 \label{syssimcaim}\,.
%% \ee
%% Contracting the last formula with $t^{\Sigma}$, and using (\ref{syssimca}), one gets a new constraint on the $\tilde{t}^{\Lambda}$s
%% \be
%% c_{\Sigma}\tilde{t}^{\Sigma}\,=\,0 \label{syssimcaim2}\,.
%% \ee
%% Assuming this constraint, the condition that the real part of the superpotential
%% vanish does not impose further constraints on the 
%% $\tilde{t}$ fields. But the two constraints (\ref{syssimca}) and 
%% (\ref{syssimcaim2}) are not sufficient to fix the  moduli since there remains a
%does not feature the fields $\tilde{t}^{\Sigma}$.
%Requiring $W=0$ also imposes no further constraints on the fields. Therefore (\ref{syssimca}) is the only constraints on the fields $\tilde{t}^{\Sigma}$.
%Since we do not want to consider solutions in which any of the 
%$\tilde{t}^{\Sigma}$ fields vanish, we must take  the determinant of $b$ to be zero. 
%But this implies that the linear system (\ref{syssimca}) has infinite solutions, so we do not fix all the moduli.
%Note that there is a further 
%% flat direction  corresponding to a rescaling of all the fields by the same number.

%%%%%%%%%%%%%%%%%%%%%%%%%%%%%%%%%%%%%%%%%%%%%%%%%%%%%%%%%%%%%%%%%%%%%%%%%%%%%%%%%%%%%%%%%%%%%%%%%%%%%%%%%
\subsubsection{No SUSY Minkowski for IIA with all moduli stabilised if $\mathrm{dim}\Delta^{2}_{+} \le \mathrm{dim}\Delta^{od}_{+}-1$}
\label{sec:nominkkcs}
%%%%%%%%%%%%%%%%%%%%%%%%%%%%%%%%%%%%%%%%%%%%%%%%%%%%%%%%%%%%%%%%%%%%%%%%%%%%%%%%%%%%%%%%%%%%%%%%%%%%%%%%%%

Recall that the number of complex structure moduli in the compactification is given by $\mathrm{dim}\Delta^{od}_{+}-1$, and the number of 
K\"ahler moduli is given by $\mathrm{dim}\Delta^{2}_{+}$. Then the no-go theorem states that me must have more K\"ahler moduli than complex structure moduli.

Consider the case $\mathrm{dim}\Delta^{2}_{+} \le \mathrm{dim}\Delta^{od}_{+}-1$. We allow $m^a \neq 0$ which implies that (\ref{conimpart})
 can be solved without imposing that all the cubic terms in the superpotential vanish.
Consider the equations $\partial_{T^a} W=0$, that depend on the complex structure moduli.
These impose the following system of equations 
\be
e_{a}^{\;\;k}u_k = f_a \label{syseq} \;,
\ee
where $f^a$ are expressions that depend on the value of the K\"ahler moduli and the dilaton  at the
minimum. 
Since the  complex structure moduli $u_k$ appear at most linearly in the superpotential,
 the conditions  $\partial_{U_k} W=0$ do not depend on these fields. Also the vanishing of the imaginary part of the superpotential adds no 
extra constraints on the $u_k$s, and the real part of the superpotential depends on the same linear combination $e_{a}^{\;\;k}u_k$ as the constraint 
(\ref{syseq}). This means that only the combination $e_{a}^{\;\;k}u_k$ of the complex structure moduli is constrained. 

Now consider the real part of $\partial_{U_k} W=0$. This imposes that the K\"ahler moduli must satisfy the following system of equations
\be 
t^a e_a^{\;\;k}=0 \label{syseq1}\;.
\ee
This implies that, since we cannot accept values $t^a=0$, the matrix $e_{a}^{\;\;k}$ can not have maximal rank. Therefore there is always at least one 
combination of the fields $u_k$ that remains unconstrained. Again, since this
direction represents the dimension of a certain cycle it can not serve as the
QCD axion and thus leaving it unfixed is undesirable. 

%%%%%%%%%%%%%%%%%%%%%%%%%%%%%%%%%%%%%%%%%%%%%%%%%%%%%%%%%%%%%%%%%%%%%%%%%%%%%%%%%%%%%%%%%%%%%%%%%%%%%%%%%
\subsection{Supersymmetric Minkowski vacua}
\label{sec:susyminkvac}
%%%%%%%%%%%%%%%%%%%%%%%%%%%%%%%%%%%%%%%%%%%%%%%%%%%%%%%%%%%%%%%%%%%%%%%%%%%%%%%%%%%%%%%%%%%%%%%%%%%%%%%%%%

In this section we consider compactifications leading to Minkowski vacua. 
We consider some tractable cases that we can solve explicitly, although 
 in section \ref{sec:genvac} we show that the
 important features of the simpler cases generalise to a larger number of moduli.
In section \ref{sec:onekahler} we study the case with a single K\"ahler modulus and no complex-structure moduli. 
From a supergravity point of view this case only really makes sense on the IIA side since the IIB equivalent would 
have no K\"ahler moduli. From a string theory point of view however this is fine and Minkowski vacua for such a case were constructed in \cite{Becker:2006ks}. 
These may be mirrors to the examples we study with no complex structure moduli on the IIA side.
 In section \ref{sec:twokahler} we consider the case with a complex structure modulus, which then 
requires (at least) two K\"ahler moduli. In section \ref{sec:threekahler} we study the case with three K\"ahler moduli and no complex structure moduli, which is the closest we can get to an explicit example. Finally in section \ref{sec:genvac} we make some statements regarding the general case.
We mostly work in IIA notation, although it is clear, at least for the cases with complex structure moduli, that these vacua are valid also in the IIB case under the mirror map outlined in section \ref{sec:mirmap}. 

%%%%%%%%%%%%%%%%%%%%%%%%%%%%%%%%%%%%%%%%%%%%%%%%%%%%%%%%%%%%%%%%%%%%%%%%%%%%%%%%%%%%%%%%%%%%%%%%%%%%%%%%%
\subsubsection{One K\"ahler modulus and no complex structure moduli}
\label{sec:onekahler}
%%%%%%%%%%%%%%%%%%%%%%%%%%%%%%%%%%%%%%%%%%%%%%%%%%%%%%%%%%%%%%%%%%%%%%%%%%%%%%%%%%%%%%%%%%%%%%%%%%%%%%%%%%

The index ranges for this case are simply $a=1$ and $k=\{ \o \}$. The superpotential (\ref{wiiahf}) simplifies to 
\begin{eqnarray}
W&=&\frac{i}{6}k m^{0} \, \left(T^{1}\right)^3
+ \frac12 k p^{1} \left( T^{1}\right)^2 
- i q_1 \,T^1 \nonumber\\ 
&-& \frac{i}{2} m^{1} k S  \left( T^{1} \right)^2  + e_{1}  S  T^{1} + i h_{0} S + e_{0} \label{wk1} \;,
\end{eqnarray}
where $k$ is the single intersection number $k=K_{111}$.
The tadpole conditions (\ref{iiatad2}) reduce to the single condition
\be
e_1 p^1 + m^0 h_0 + q_1 m^1 = -Q_0\;. 
\ee
As discussed in section \ref{sec:fluxax} we do not include a possible axion contribution.
We solve this tadpole equation by taking
\be
q_1 = -\frac{1}{m^1} \left( e_1 p^1 + m^0 h_0 + Q_0 \right) \label{1kq1}\;.
\ee
This eliminates $q_1$ from further equations and guarantees that the tadpole is solved. However it also places the constraint on the other fluxes that 
there is a solution to (\ref{1kq1}) with $q_1$ integer. 

We proceed to solve the equations (\ref{susyminkeq}) for the superpotential (\ref{wk1}).
Imposing $\partial_S W=0$ gives 
\ba
b^1 &=& -\frac{e_1}{m^1 k} \;, \\
\left(t^1\right)^2 &=& \frac{2 h_0}{m^1 k} - \frac{e_1^2}{\left(m^1 k \right)^2} \label{1kt} \;.
\ea
Since we would like vacua with vanishing axion we henceforth set $e_1=0$ thereby justifying not including it in the tadpole condition.
The condition $\partial_{T^1}W=0$ gives 
\begin{eqnarray}
\sigma &=& -\frac{p_1}{m^1}\;,  \\
s&=& \frac{1}{k \left(m^{1}\right)^2 t^{1}} \,\left( 2 m^0 h_0 + Q_0 \right) \;.\label{1ks}
\end{eqnarray}
We now wish to impose the condition $W=0$. We begin by solving for a vanishing imaginary part and return to the real part later. 
The combination
\be
\im{\left( t^1 \partial_{T^1} W + s \partial_{S} W - W \right)}\,=\,0 \;, 
\ee
becomes the constraint
\be
3 m^1 s = m^0 t^1 \;,
\ee
which can be solved by taking
\be
h_0 = -\frac{3 Q_0}{4 m^0} \;. \label{1kh0}
\ee
Again we have the constraint on the fluxes that there is a solution for $h_0$ integer. Substituting this into (\ref{1ks}) and (\ref{1kt}) we recover 
\ba
s  &=& \frac{1}{m^1}\sqrt{\frac{-Q_0 m^0}{6 k m^1 }} \;, \\
t^1  &=& \sqrt{\frac{-3Q_0}{2 m^0 m^1 k}} \;.
\ea
Now we see that the value of the K\"ahler modulus is capped by the orientifold charge. Although we can go to arbitrarily weak coupling this decreases the 
value for $t^1$ such that the product $s t^1$ is constant. 
Note also that if $Q_0 > 0$ there are no solutions. We therefore satisfy the no-go theorem of \cite{Grana:2006kf} which states that 
all the Minkowski vacua must have orientifolds present.
We also need to check that the real part of the superpotential vanishes. This can always be done by choosing $e_0$ appropriately, with the constraint on the 
fluxes that it should be integer. For this case the solution is 
\be
e_0 =\frac{3 p^1 Q_0}{4 m_0 m_1 } \;. \label{1ke0int}
\ee
We now need to check that we can choose the fluxes such that they are all integer. We first note that the flux $p^1$ only features in the value of the axion 
$\sigma$. Then we are free to choose it as we like without changing the values of the geometric moduli. It is clear that it can be chosen such that $q_1$ and $e_0$, as written in (\ref{1kq1}) and (\ref{1ke0int}), are integers. The flux $h_0$, as in (\ref{1kh0}), can be made integer by taking $m^0$ to be 1 or 3, 
since $Q_0$ must be a multiple of 4. We therefore see that we have a fully consistent solution, which for enough orientifolds, can be at large values 
for the moduli. 

We do not perform any analytic vacua counting
  since there are too many constraints to be satisfied by the fluxes. 
We note that, although $p^1$ only features in the axion value, the number of vacua is not infinite since in counting vacua we should gauge fix 
the integer axionic shift symmetry of the superpotential  
\be
\sigma \rightarrow \sigma + 1 \;, \;\; p^1 \rightarrow p^1 - m^1 \;, \;\; q_1 \rightarrow q_1 + 1 \;, \;\; e_0 \rightarrow e_0 + h_0 \;,
\ee
by taking 
\be
0 \le p^1 < |m^1| \;.
\ee

There are a number of known nearly-K\"ahler manifolds that have $SU(3)$-structure which have a single K\"ahler modulus and no complex-structure moduli, see 
\cite{Behrndt:2004mj} for a list. We would require a non-geometric deformation of these manifolds to reach the vacua we have found in this section. See section \ref{sec:threekahler} for a similar point for the case with three K\"ahler moduli. 

To summarise we find Minkowski vacua with all the moduli stabilised and tadpole conditions satisfied. The 
values of the moduli and dilaton are capped by the orientifolds charge. 

%%%%%%%%%%%%%%%%%%%%%%%%%%%%%%%%%%%%%%%%%%%%%%%%%%%%%%%%%%%%%%%%%%%%%%%%%%%%%%%%%%%%%%%%%%%%%%%%%%%%%%%%%
\subsubsection{Two K\"ahler and one complex structure moduli}
\label{sec:twokahler}
%%%%%%%%%%%%%%%%%%%%%%%%%%%%%%%%%%%%%%%%%%%%%%%%%%%%%%%%%%%%%%%%%%%%%%%%%%%%%%%%%%%%%%%%%%%%%%%%%%%%%%%%%%

We now study the two K\"ahler moduli and one complex structure modulus case. The index ranges are therefore $a=1,2$ and $k=1$.
We have the superpotential
\begin{eqnarray}
W&=&\frac{i}{6} m^{0} k_0 \left(T^{1}\right)^3 + \frac{i}{2} m^{0} k_1 T^2 \left(T^{1}\right)^2 + \frac{i}{2} m^{0} k_2 T^1 \left(T^{2}\right)^2 +\frac{i}{6} m^{0} k_3 \left(T^{2}\right)^3 \nn \\
 &+& \frac12 k_0 p^{1} \left( T^{1}\right)^2 + \frac12 k_2 p^{1} \left( T^{2}\right)^2 + \frac12 k_1 p^{2} \left( T^{1}\right)^2 + \frac12 k_3 p^{2} \left( T^{2}\right)^2 + k_2 p^{2} T^1 T^2 \nn \\
 &+& k_1 p^{1}  T^1 T^2  - i q^{1} \,T^{1} -i q_2 T^2 + e_0 \nonumber\\ 
&-& \frac{i}{2} k_2 m^{1} S \left( T^2 \right)^2 - \frac{i}{2} k_1 m^{2} S \left( T^1 \right)^2 - \frac{i}{2} k_3 m^{2} S \left( T^2 \right)^2 - i k_2 m^{2} S T^1 T^2 - i k_1 m^{1} S T^1 T^2 \nn \\ 
&-& \frac{i}{2} k_0 m^{1} S \left( T^1 \right)^2 +   e_1  S   T^1 + e_2 S T^2 + e_3 U T^1 + e_4 U T^2 + i h_0 S + i h^1 U \;,
 \label{W2k}
\end{eqnarray}
where we denote $e_{1}^{\;\;1}=e_3,e_{2}^{\;\;1}=e_4$.
The intersection numbers are denoted as
\ba
k_0 &=& K_{111} \;,\nn \\
k_1 &=& K_{112} \;,\nn \\
k_2 &=& K_{122} \;,\nn \\
k_3 &=& K_{222} \;.
\ea
The tadpole conditions read
\ba
p^1 e_1 +  p^2 e_2 + m^0 h_0 + q_1 m^1 +q_2 m^2 &=& -Q_0 \label{2ktp1} \;,\\
p^1 e_3 + p^2 e_4 + m^0 h^1 &=& - Q^1 \label{2ktp2} \;,\\
m^1 e_3 + m^2 e_4 &=& 0 \label{2ktp4} \;.
\ea
We solve the tadpole conditions by fixing $q_1$, $h_1$ and $m_2$ in terms of the other fluxes whilst keeping in mind that we have the 
constraints that they should be integer. 

We now go on to solve the supersymmetry variations. The solutions for the axions read
\ba
b^1 &=& \frac{ m^0 { {e_4}}^2 \left( {e_2}\, {e_3}\; -  {e_1}\,e_4\, \right)- 
    \left( { {e_4}}^2\, {k_1} - 2\, {e_3}\, {e_4}\, {k_2} + { {e_3}}^2\, {k_3} \right) \,
      {m^1}\,\left(  {e_3}\, {p^1} +  {e_4}\, {p^2} +  {Q^1} \right) }{
    \left( { {e_4}}^3\, {k_0} - 3\, {e_3}\,{ {e_4}}^2\, {k_1} + 
      3\,{ {e_3}}^2\, {e_4}\, {k_2} - { {e_3}}^3\, {k_3} \right) \, {m^0}\, {m^1}}\nn \;,\\
b^2 &=& \frac{- m^0 e_3 e_4 \left(  e_2\,e_3 - e_1 \; e_4  \right)  + 
    \left( { {e_4}}^2\, {k_0} - 2\, {e_3}\, {e_4}\, {k_1} +
{ {e_3}}^2\, {k_2} \right) \,
      {m^1}\,\left(  {e_3}\, {p^1} +  {e_4}\, {p^2} +  {Q^1} \right) }{
    \left( { {e_4}}^3\, {k_0} - 3\, {e_3}\,{ {e_4}}^2\, {k_1} + 
      3\,{ {e_3}}^2\, {e_4}\, {k_2} - { {e_3}}^3\, {k_3} \right)
\, {m^0}\, {m^1}} \nn \;.
\ea
As discussed in section \ref{sec:fluxax} we impose the conditions that the combinations $e_{a} b^a$ and $ e_a^{\;\;k} b^a$ vanish giving the constraints
\ba
e_2 e_3 &=& e_1 e_4 \;, \nn \\
{e_3}\, {p^1} +  {e_4}\, {p^2} +  {Q^1} &=& h^1 = 0 \;.
\ea
We can solve these conditions by eliminating $e_1$ and $p^1$.
The moduli values in the vacuum then read
\ba
t^1 &=& e_4 \sqrt{\frac32}\sqrt{\frac{ -e_4 Q_0 + e_2 Q^1}{m^0 m^1 F_1}} \;,\nn \\
t^2 &=& -\frac{e_3}{e_4}\; t^1 \;,\nn \\
s &=& \frac{m^0}{3 m^1}\; t^1 \;,\nn \\
u &=& \frac{3 m^1 Q^1 F_2 - e_2 m^0 F_1}{3 e_4 m^1 F_1} \;t^1\;, \label{minksol2k}
\ea
where we define the quantities
\ba
F_1 &\equiv & \left( e_4 \right)^3 k_0 - 3 e_3 \left(e_4\right)^2 k_1 + 3 e_4 \left(e_3 \right)^2 k_2 - \left( e_3 \right)^3 k_3  \;, \\
F_2 &\equiv & - \left( e_4 \right)^2 \left( k_1 \right)^2 + \left( e_4 \right)^2 k_0 k_2 + e_3 e_4 k_1 k_2 - e_3 e_4 k_0 k_3 + \left( e_3 \right)^2 k_1 k_3 - \left( e_3 \right)^2 \left( k_2 \right)^2 \;. \nn
\ea
We have eliminated $h_0$ by requiring that the imaginary part of the superpotential vanishes. 
Its explicit value is given in the appendix along with the values of the axions. Similarly the real part of the superpotential fixes the flux $e_0$.
The solution (\ref{minksol2k}) is a Minkowski vacuum with all the moduli stabilised. The values of the fluxes can be chosen so that all the moduli 
are positive and large. We later study the number of such vacua, but for now we outline a family of solutions where the moduli can be made parametrically 
large. Consider setting the fluxes $e_4=-e_3=m^0=m^1=1$ and taking $e_2$ large. Then the solution for the moduli reads
\ba
t^1 = t^2 = 3 s &\sim& \sqrt{\frac{3}{2 F_1}}\sqrt{e_2 Q^1} \;,\nn \\
u &\sim& -\frac13 e_2 \;t^1\;. \label{minksimpsol2k}
\ea
Then if we take $Q^1<0$ and $e_2<0$ we have a parametrically controlled family of solutions where we can reach arbitrarily large values for the moduli. 
In terms of the size moduli and string coupling\footnote{In the type IIA case the ten-dimensional dilaton scales as 
$e^{-\hat{\phi}} \sim \left(\frac{su^3}{t^6}\right)^{\frac14}$.} this solution reads
\ba
IIA\;&:& \;\; t^1,\;t^2 \sim \left|e_2\right|^{\frac12}\;\;,\;\; e^{-\hat{\phi}} \sim \left|e_2\right|^{\frac12} 
\;,\;\; u \sim \left| e_2 \right|^{\frac32} \;, \nn \\
IIB\;&:& \;\; u^1,\;u^2 \sim \left|e_2\right|^{\frac12}\;\;,\;\; e^{-\hat{\phi}} \sim \left|e_2\right|^{\frac12} \;,\;\; 
t \sim \left| e_2 \right|^{\frac32} \;.
\ea

At this point it is worth noting how the parametrically controlled vacua satisfy the no-go theorem of \cite{Grana:2006kf}. 
Although it is clear that we require orientifolds wrapped over $\alpha_1$, since $Q^1<0$, 
this need not be the case for $Q_0$ since it can be positive. At first this leads to an apparent contradiction since we can take $Q_0$ large and positive and $Q^1$ small and negative (which is fine as long as $e_2$ is large and negative) so that if we sum over the total charges present 
the result is an overall positive charge. The resolution of this is to note that the no-go theorem states that 
\be
\int_{Y}{\left< \im \Pi_c^{od}, \delta_{source}\right>} = Q_0 s + Q^1 u \sim |e_2|^{\frac12}\left( Q_0 + Q^1 |e_2| \right) < 0 \;.
\ee 
Hence we see that the charge $Q^1$ is weighted by an extra factor of $|e_2|$ thereby satisfying the no-go theorem.

We still have the requirement on the solutions that the tadpoles and other consistency equations are solved for integer 
values of the fluxes. Because there are may of these constraints it is difficult to perform an analytical estimate 
of the number of vacua within a flux range. Also since we are not restricted to a particular manifold we have a large number of free parameters. A full analysis of the number of vacua is therefore beyond the scope of this paper. 
Instead we present a simple analysis, which is intended to give an idea for the numbers. We consider the intersection numbers\footnote{The intersection numbers are those of the complete intersection Calabi-Yau $\left(  \begin{array}{cc} 2 \;|& 3  \\ 3 \;|& 4   \end{array} \right)$.}
\be
K_{111} = 0\;, \;\;K_{112} = 0\;, \;\;K_{221} = 4\;, \;\;K_{222} = 2\;. \label{examint}
\ee 
We count the number of Minkowski vacua that have all the moduli larger than some value $t^1,t^2,u,s \ge X_{\mathrm{min}}$ where all the tadpoles vanish and the fluxes take integer values. We also fix the the orientifold/D-branes charges to be $Q_0=Q^1=-32$. We then scan over the flux parameters $e_4,e_3,e_2,m^0,m^1$ within an integer range $-M,...,M$. We also have the free fluxes $p^2$ and $q_2$. However we restrict the values of these fluxes in order to fix left over integer axionic symmetries so as not to overcount the vacua drastically. The symmetries are 
\be
\sigma \rightarrow \sigma + 1 \;,
\ee
which can be absorbed into a flux redefinition 
\be
e_0 \rightarrow e_0 + h_0 \;,\;\; p^a \rightarrow p^a - m^a \;,\;\; q_a \rightarrow q_a + e_a\;. \label{axshex}
\ee
We fix this by constraining 
\be
0 \le p^2 < |m^2| = - \frac{m^1 e_3}{e_4} \;.
\ee
Similarly we fix the shifts of $\nu$ by constraining 
\be
0 \le q_2 < |e_4| \;.
\ee
With these constraints the number of vacua are presented in table \ref{tab:vac}. Note that the vacua are  relatively 
sparse. This is primarily because the constraint on $e_0$ being integer which is difficult to satisfy.
\begin{table}
\center
 \begin{tabular}{||c|c|c|c|c|c||} \hline
 $M$/$X_{\mathrm{min}}$ & 1 & 2 & 3 & 4 & 5\\ \hline
 $10$ & 33 & 0 & 0 & 0 & 0\\ \hline
 $20$ & 206 & 5 & 0 & 0 & 0\\ \hline
 $30$ & 481 & 33 & 0 & 0 & 0\\ \hline
 $40$ & 898 & 94 & 5 & 0 & 0\\ \hline
 $50$ & 1525 & 250 & 48 & 12 & 6\\ \hline
\end{tabular}
\caption{Table showing the number of vacua with all the moduli larger than $X_{\mathrm{min}}$ as a function of the flux 
parameters range $M$ for the intersection number choices (\ref{examint}) and the orientifold charges $Q_0=Q_1=-32$.}
\label{tab:vac}
\end{table}   

To summarise, we find that in the case of two K\"ahler moduli and one complex structure modulus there is a parametrically controlled family of Minkowski vacua 
with arbitrary large values for the moduli. In section \ref{sec:genvac} we show that this situation is generic as long as we have at least one complex structure modulus.

%%%%%%%%%%%%%%%%%%%%%%%%%%%%%%%%%%%%%%%%%%%%%%%%%%%%%%%%%%%%%%%%%%%%%%%%%%%%%%%%%%%%%%%%%%%%%%%%%%%%%%%%%
\subsubsection{Three K\"ahler moduli and no complex structure moduli}
\label{sec:threekahler}
%%%%%%%%%%%%%%%%%%%%%%%%%%%%%%%%%%%%%%%%%%%%%%%%%%%%%%%%%%%%%%%%%%%%%%%%%%%%%%%%%%%%%%%%%%%%%%%%%%%%%%%%%

In this section we consider the case of three K\"ahler moduli and no complex moduli. The main reason for studying this is 
that $SU(3)$ structure examples of these manifolds are known. For example the tori in \cite{DeWolfe:2005uu} and the
coset manifold in \cite{House:2005yc}. We therefore require non-geometric deformations of these manifolds in order to reach 
the type of superpotentials we are studying. 
These could be obtained by T-dualising the torus with H-flux along two
directions and then modding out by a $Z_3 \times Z_3$ orbifold symmetry.
We leave a solid construction of such manifolds for future work and go on to study the superpotential. Since both the known examples are parallelisable manifolds on which the two-forms are constructed as a product of one-forms, the only non-vanishing intersection number 
is the one with no repeated indices. We make the simplifying assumption that this is also the case at hand and take the only non-vanishing intersection
\be
K_{123} = 1 \;.
\ee
With this the superpotential reads
\ba
W &=& im^0 T^1 T^2 T^3 + p^1 T^2 T^3 + p^2 T^1 T^2 + p^3 T^1 T^2 - iq_1 T^1- iq_2 T^2 - iq_3 T^3 + e_0 \nn \\ 
&\;& -iS m^1 T^2 T^3 -iS m^2 T^1 T^3 -iS m^3 T^1 T^2 + S e_1 T^1 + S e_2 T^2 + S e_3 T^3 + ih_0 S \;.
\ea 
Since there are no new features in this type of superpotential rather than solving it generally we look for a 
particular solution. It is easy to find a solution where the fields are all equal and the fluxes with the varying indices are set equal. 
Then solving the supersymmetry equations is equivalent to solving the one K\"ahler modulus case as 
in section \ref{sec:onekahler}. The solution reads 
\ba
b^1 &=& b^2 = b^3 = 0 \;, \nn \\
t^1 &=& t^2 = t^3 = \sqrt{\frac{-Q_0}{4m^0m}} \;, \nn \\
\sigma &=& -\frac{p}{m} \;, \nn \\
s &=& \frac{m^0}{3m}\sqrt{\frac{-Q_0}{4m^0m}} \;,
\ea
where we solved the tadpoles by setting 
\be
q = -\frac{1}{3m}\left( m^0 h_0 + Q_0 \right) \;.
\ee

%%%%%%%%%%%%%%%%%%%%%%%%%%%%%%%%%%%%%%%%%%%%%%%%%%%%%%%%%%%%%%%%%%%%%%%%%%%%%%%%%%%%%%%%%%%%%%%%%%%%%%%%%
\subsubsection{The general case}
\label{sec:genvac}
%%%%%%%%%%%%%%%%%%%%%%%%%%%%%%%%%%%%%%%%%%%%%%%%%%%%%%%%%%%%%%%%%%%%%%%%%%%%%%%%%%%%%%%%%%%%%%%%%%%%%%%%%

In this section we discuss the general case where the number of moduli fields is arbitrary up to the constraint 
$\mathrm{dim}\Delta^{2}_{+} > \mathrm{dim}\Delta^{od}_{+}-1$. Then
  equation (\ref{syseq}) means that if the matrix 
$e_a^{\;\;k}$ does not have maximal rank we can not stabilise all the complex structure moduli.  
We set out to show that if $e_a^{\;\;k}$ does have maximal rank then there are always solutions with all the moduli stabilised. 
Further we show that, as long as there is at least one complex structure modulus, these solutions are ones where the moduli can be parametrically 
taken to arbitrary large values. We do this by outlining a particular class of solutions, rather than finding the most general vacuum. 
We do not impose the integer constraints on the fluxes and assume that, by performing scans
 like the one in section \ref{sec:twokahler}, solutions can be found where all the fluxes are integers.  

Before proceeding it helps to introduce some notation for the index ranges of the fluxes and moduli. Let there be $n=\mathrm{dim}\Delta^{2}_{+}$ 
K\"ahler moduli with the index $a$ running from $1,...,n$. Then we introduce the indices $\tilde{a}$ and $\bar{a}$ which have the ranges 
$\tilde{a}=1,...,p$ and $\bar{a}=p+1,...,n$, where $p$ is the number of complex structure moduli $p=\mathrm{dim}\Delta^{od}_{+}-1$.

We now go through the conditions for a supersymmetric Minkowski vacuum and indicate the degrees of freedom fixed by each condition. 
We look for a solution where all the NS axions vanish $b^a=0$. This is
sufficient to have no ambiguity in the tadpole conditions. 

We start with the condition on the superpotential vanishing. The imaginary part of this condition reads 
\be
\label{nogos}
\frac{m^0}{3} K_{a b c} t^a t^b t^c\,=\,s K_{a b c} m^a t^b t^c \;.
\ee  
We solve this condition by fixing the value of the dilaton $s$. We also impose positive intersection numbers and positive values for for the fluxes 
$m^0$ and $m^a$ so that if the $t^a$ are positive so is the dilaton. We also see that the dilaton scales like $t^a$.
The real part of the superpotential imposes a condition on $e_0$ which fixes its value.
Now, consider the real part of the condition $\partial_S W=0$ which gives 
\be
e_{\tilde{a}} t^{\tilde{a}} + e_{\bar{a}} t^{\bar{a}} = 0\,. \label{ebar1}
\ee
We solve this condition by fixing one of the fluxes $e_{\bar{a}}$. 
The imaginary part of the derivative gives 
\be
\label{conh0}
h_0=\frac12 K_{a b c} m^a t^b t^c\,,
\ee
which we solve by fixing $h_0$.
The imaginary parts of the conditions $\partial_{U_k} W=0$ impose that $h^k=0$. The real parts give the $p$ conditions
\be
e^{\;\;k}_{\tilde{a}} t^{\tilde{a}}+e^{\;\;k}_{\bar{a}} t^{\bar{a}}=0 \label{system1} \;.
\ee
Since the matrix $e^{\;\;k}_{a}$ has maximal rank, we can choose
 the square matrix $E_1 \equiv e^{\;\;k}_{\tilde{a}}$ to have
  non-vanishing determinant. Therefore, if we also define
  $E_2 \equiv e^{\;\;k}_{\bar{a}}$, we can solve the constraints (\ref{system1}) by taking
\be
t^{\tilde{a}} = - \left( {E_1}^{-1} E_2 \right)^{\tilde{a}}_{\;\;\bar{a}} t^{\bar{a}} \;.
\ee 
We now impose the conditions on the fluxes that 
$-\left( {E_1}^{-1} E_2 \right)^{\tilde{a}}_{\;\;\bar{a}}$ is positive. This does not fix the fluxes 
but does limit their possible values. It means that if the $t^{\bar{a}}$ are positive, so are $t^{\tilde{a}}$. 
For alter convenience we also impose $E_1=\left( E_1 \right)^T$ which halves the degrees of freedom. 
Let us now focus on a solution in which $t^{\bar{a}}=\alpha m^{\bar{a}}$.
The real constant $\alpha$ sets the scale of the moduli $t^a$ and $s$.
The final set of conditions come from $\partial_{T^a} W=0$. These are $n$ constraints that read 
\be
\frac{i}{2} m^0 K_{a bc} T^b T^c + K_{a bc} p^b T^c -i q^a - i K_{a bc} S m^b T^c +e_a S+ e_a^{\;\;k} U_k\,=\,0 \;.
\label{derT} 
\ee
Taking the real part of (\ref{derT}) gives 
\be
 K_{abc} p^b t^c+\sigma K_{abc}m^b t^c+s e_a + e^{\;\;k}_a u_{k}\,=\,0\label{sysu} \;.
\ee
We now contract the last expression with $t^a$ which gives an equation that fixes $\sigma$ in terms of $t^a$
\be
\sigma\,=\,-\frac{K_{abc}p^a t^b t^c}{K_{abc}m^a t^b t^c} \;.
\ee
There remain $n-1$ constraints. The first $p$ of these fix the real parts of the complex moduli
\be
u_k=\,-\left({E_1}^{-1}\right)^{\;\;\tilde{a}}_{k}\left[ e_{\tilde{a}} s +  K_{\tilde{a}bc} p^b t^c - 
K_{\tilde{a}bc}m^b t^c\frac{K_{bcd}p^b t^c t^d}{K_{bcd}m^b t^c t^d} \right]\,, \label{usol}
\ee
Later we take, $-\left({E_1}^{-1}\right)^{\;\;\tilde{a}}_{k}e_{\tilde{a}} \gg 1$, in which case 
we see that the $u_k$ are positive.
Putting the solutions (\ref{usol}) back into the remaining $(n-p-1)$ components of (\ref{sysu}), we find $(n-p-1)$ constraints
on the fluxes which we solve by fixing the remaining $e_{\bar{a}}$ fluxes (recall that 
one was fixed by (\ref{ebar1})). Similar arguments apply when taking the imaginary part of (\ref{derT}) which reads 
\be 
\frac{m^0}{6} K_{a b c}t^b t^c -q_a -\frac{s}{2} K_{a b c} m^b t^c + e_a \sigma+e^{\;\;k}_{a} \nu_k\,=\,0\label{eqfornu} \;.
\ee
The first $p$ of these conditions fix the axions $v_k$ and the remaining $(n-p)$ equations give constraints on the fluxes. 
We solve these by fixing the fluxes $q_{\bar{a}}$.

We have now solved all the supersymmetry equations and can apply the tadpole conditions to these solutions. The tadpole conditions read
\ba
m^a e_{a}^{\;\;k} &=& 0 \;. \label{iiatadme2}\\
- m^0 h_0 - p^a e_{a} - m^a q_a &=& Q_0 \label{iiatad22}\;, \\
- p^a e_{a}^{\;\;k} &=& Q^k  \;, \label{iiatad33}
\ea
The $p$ constraints (\ref{iiatadme2}) are solved by fixing $m^{\tilde{a}}$.
We now set $p^{\bar{a}}=0$ which means that the $p$ conditions of (\ref{iiatad33}) can be solved by taking 
\be
p^{\tilde{a}} = -\left( {E_1}^{-1}\right)^{\tilde{a}}_{\;\;k} Q^k \;.
\ee
We are left with the single tadpole equation (\ref{iiatad22}).
This tadpole fixes the overall scale factor $\alpha$, that is, the overall scale of the moduli.
  To see this consider contracting (\ref{eqfornu}) with $t^a$, 
and using (\ref{nogos}). One finds the relation
\be
%\label{combi}
q_a t^a\,=\,\frac{m^0}{6} K_{a b c} t^a t^b t^c\,=\,\frac{s}{2} K_{a b c} t^a m^b t^c \;.\label{qeq1}
\ee
Now we set $q^{\tilde{a}}=0$ which means that (\ref{qeq1}) gives 
\be
q_{\bar{a}} m^{\bar{a}} = \alpha^2 H_1\;,
\ee
where $H_1$ is a real number that depends on $K_{abc}$, $e_{a}^{\;\;k}$, $m^{\bar{a}}$ and $m^0$. Also (\ref{conh0}) gives
\be
m^0 h_0 = \alpha^2 H_2 \;,
\ee
where $H_2$ is a real number which depends on $K_{abc}$, $e_{a}^{\;\;k}$, $m^{a}$ and $m^0$. Therefore using (\ref{iiatad22}) we can write 
\be
\alpha^{2}\,=\, \frac{-Q_0+e_{\tilde{a}}\left({E_1}^{-1}\right)^{\tilde{a}}_{\;\;k}\,Q^{k}}{H_2 + H_1} \;.
\ee
But this last formula ensures that, by taking $Q^k$ negative and $-\left({E_1}^{-1}\right)^{\;\;\tilde{a}}_{k}e_{\tilde{a}} \gg 1$, one can find parametrically large values for $\alpha$ and so for all the moduli. 

%%%%%%%%%%%%%%%%%%%%%%%%%%%%%%%%%%%%%%%%%%%%%%%%%%%%%%%%%%%%%%%%%%%%%%%%%%%%%%%%%%%%%%%%%%%%%%%%%%%%%%%%%
\subsection{Ten-dimensional uplifts and torsion classes}
\label{sec:10dsol}
%%%%%%%%%%%%%%%%%%%%%%%%%%%%%%%%%%%%%%%%%%%%%%%%%%%%%%%%%%%%%%%%%%%%%%%%%%%%%%%%%%%%%%%%%%%%%%%%%%%%%%%%%%

So far we have considered type II theories on non-geometric backgrounds with
$SU(3) \times SU(3)$ structure and we have studied solutions of the
four-dimensional truncations thereof. However, until now we have not studied
whether such backgrounds are consistent string backgrounds or if the solutions
obtained are indeed solutions of the full ten-dimensional string/supergravity
theories. In this section we precisely want to fill this gap.

A simple argument to show the consistency of our solutions with the
ten-dimensional picture is to note that the superpotential of
\cite{Benmachiche:2006df}, which we generalised in our paper, was derived from
the reduction of the fermionic action. Therefore, the solution to the
four-dimensional supersymmetry equations derived from such a superpotential
will solve the ten-dimensional supersymmetry variations as well. 

For the case at hand this can also be seen explicitely as follows.
The conditions for $N=1$ Minkowski vacua, which spelled out in
\cite{Grana:2005sn} (see also \cite{Grana:2004bg}), 
require that the internal manifold is a twisted generalised Calabi--Yau
manifold
%\footnote{This 
%  is yet different to the generalised Calabi--Yau condition of \cite{GCY} for
%  which the pure spinor is closed.} 
 defined by the condition
\be
d_H \Pi^{ev} = 0 \;, \label{gency2} 
\ee
provided we are in a regime where the dilaton and the warp factor are constant
over the internal manifold.

As we argued before in the paper, for the non-geometric backgrounds we
consider one has to replace the twisted exterior derivative $d_H$ by the 
covariant derivative of \eqref{tcd}. Therefore, for our case the condition for
$N=1$ Minkowski vacua reads
\be
\D \Pi^{ev} = 0 \;. \label{gency3} 
\ee
It is possible to see then that the above equation is equivalent to the  
supersymmetry conditions corresponding to the superpotential (\ref{wiiahf})
along the directions $S$ and $U^k$. This implies that in a supersymmetric
Minkowski vacuum the moduli take values such that the manifold is twisted
(non-geometric) generalised Calabi-Yau as required by
  the ten-dimensional
analysis. 

\section{Summary}
%%%%%%%%%%%%%%%%%%%%%%%%%%%%%%%%%%%%%%%%%%%%%%%%%%%%%%%%%%%%%%%%%%%%%%%%%%%%%%%%%%%%%%%%%%%%%%%%%%%%%%%%%%

In this paper we studied compactifications of type IIA string theory on manifolds with $SU(3) \times SU(3)$ structure in the presence of $O6$-planes and its mirror compactification of type IIB on manifolds with $SU(3) \times SU(3)$ structure in the presence of $O3$-planes. We argued that generalised geometry 
provides us with the tools needed to treat these non-geometric compactifications in a geometric sense. By introducing a covariant derivative for T-dualities 
we were able to derive the four-dimensional superpotential.
We showed that, in the presence of non-geometric fluxes, if the number (in IIA notation) of K\"ahler moduli is larger than the number of complex structure moduli, the theory contains supersymmetric Minkowski vacua with all the moduli stabilised in a a perturbative regime. We find that if there are no complex structure moduli then the value of the moduli in the vacuum is capped by the orientifold charge. In the presence of complex structure moduli however there are parametrically controlled vacua. 

The possible extensions to this work are numerous, and we hope that the systematic construction of Minkowski vacua presented in this paper will help 
eliminate the need to uplift the usual anti deSitter vacua to Minkowski, and all the problems associated with this mechanism such as 
fine tuning and high scale supersymmetry breaking.

%%%%%%%%%%%%%%%%%%%%%%%%%%%%%%%%%%%%%%%%%%%%%%%%%%%%%%%%%%%%%%%%%%%%%%%%%%%%%%%%%%%%%%%%%%%%%%%%%%%%%%%%%
\vspace{1.5cm}

{\Large Acknowledgments}
\newline
%%%%%%%%%%%%%%%%%%%%%%%%%%%%%%%%%%%%%%%%%%%%%%%%%%%%%%%%%%%%%%%%%%%%%%%%%%%%%%%%%%%%%%%%%%%%%%%%%%%%%%%%%

The authors would like to thank Gil Cavalcanti, James Gray, Thomas Grimm, Andre Lukas and David Skinner for useful discussions.

EP is supported by a PPARC Postdoctoral Fellowship. 
GT and AM were partially supported by the EC $6^{th}$ Framework Programme
MRTN-CT-2004-503369, and  by the EU 6th Framework Marie Curie        
Research and Training network " (MRTN-CT-2006-035863) ``UniverseNet''.
AM was also supported by the European project MRTN-CT-2004-005104
"ForcesUniverse" and SFB-Transregio 33 "The Dark Universe" by Deutsche
Forschungsgemeinschaft (DFG).

%%%%%%%%%%%%%%%%%%%%%%%%%%%%%%%%%%%%%%%%%%%%%%%%%%%%%%%%%%%%%%%%%%%%%%%%%%%%%%%%%%%%%%%%%%%%%%%%%%%%%%%%%
\vspace{1cm}
%\newpage
\appendix
\noindent\textbf{\LARGE Appendix}

%%%%%%%%%%%%%%%%%%%%%%%%%%%%%%%%%%%%%%%%%%%%%%%%%%%%%%%%%%%%%%%%%%%%%%%%%%%%%%%%%%%%%%%%%%%%%%%%%%%%%%%%%

%%%%%%%%%%%%%%%%%%%%%%%%%%%%%%%%%%%%%%%%%%%%%%%%%%%%%%%%%%%%%%%%%%%%%%%%%%%%%%%%%%%%%%%%%%%%%%%%%%%%%%%%%
\section{Full solution with two K\"ahler moduli}
%%%%%%%%%%%%%%%%%%%%%%%%%%%%%%%%%%%%%%%%%%%%%%%%%%%%%%%%%%%%%%%%%%%%%%%%%%%%%%%%%%%%%%%%%%%%%%%%%%%%%%%%%%

In this appendix we include for completeness the rest of the two K\"ahler moduli case studied in section \ref{sec:twokahler}. We also 
include a set of flux parameters as an explicit solution. The axions read
\ba
b^1 &=& b^2 = 0 \;, \nn \\
\sigma &=& \frac{e_4}{e_3m^1} \left( p^2 + \frac{Q_1\left({e_4}^2 k_0 - 2 e_3 e_4 k_1 + {e_3}^2 k_2 \right)}{F_1} \right)\;, \nn \\
\nu &=& \frac{-1}{4e_3e_4F_1m^1}\left[ 4 e_2 e_4 F_1 p^2 - 4e_3 F_1 m^1 q_2 - e_3 e_4 \left( {e_4}^2 k_1 - 2 e_3 e_4 k_2 + {e_3}^2 k_3 \right) Q_0 
\right. \nn \\  &\;& \;\;\;\;\;\;\;\;\;\;\;\;\;\;\;\;\;\;\;\;\; \left.  + e_2 \left( 4 {e_4}^3 k_0 
- 7 e_3 {e_4}^2 k_1 + 2 {e_3}^2 e_4 k_2 + {e_3}^3 k_3 \right)Q_1 \right]\;.
\ea
The fixed fluxes are
\ba
m^2 &=& - m^1 \frac{e_3}{e_4} \nn \;, \\
e_1 &=& \frac{e_2 e_3}{e_4} \;, \nn \\
p^1 &=& -\frac{e_4 p^2 + Q_1}{e_3} \;, \nn \\
q_1 &=& \frac{1}{m^1} \left( - Q_0 + \frac{q_2 m^1 e_3}{e_4} - p^1 e_1 - p^2 e_2 - m^0 h_0 \right) \nn \;, \\
h_0 &=& \frac{3\left(-e_4 Q_0 + e_2 Q_1 \right)}{4 e_4 m^0} \;,\nn \\
h_1 &=& 0 \;, \nn \\
e_0 &=& - \frac{3}{4e_3{F_1}^2m^0m^1} \left[ \left({e_4}^3 k_0 - 4 e_3 {e_4}^2 k_1 + 4{e_3}^2 e_4 k_2 - {e_3}^3 k_3 \right)\left(e_4 Q_0 - e_2 Q_1 \right) \right. \nn \\ 
&\;& \;\;\;\;\;\;\;\;\;\;\;\;\;\;\;\;\;\;\;\;\;\;\; \left.\left( p^2 F_1 + Q_1 \left( {e_4}^2 k_0 - 2 e_3 e_4 k_1 + {e_3}^2 k_2 \right)\right)\right] \;.
\ea
An explicit example of a solution for the intersection numbers (\ref{examint}) is presented in table \ref{tab:sol}
\begin{table}[h]
\center
 \begin{tabular}{||c|c|c|c|c|c|c|c|c|c|c|c|c|c|c|c|c|c||} \hline
 $v^1$ & $v^2$ & $s$ & $u$ & $m^0$ & $m^1$ & $m^2$ & $e_0$ & $e_1$ & $e_2$ & $e_3$ & $e_4$ & $p^1$ & $p^2$ & $q_1$ & $q_2$ & $h_0$ & $h_1$ \\ \hline
 4.6 & 4.6 & 6.1 & 154 & 4 & 1 & 1 & 432 & 94 & -94 & -4 & 4 & -8 & 0 & 196 & 0 & 147 & 0  \\ \hline
\end{tabular}
\caption{Table showing values of flux parameters and moduli values for an explicit solution.}
\label{tab:sol}
\end{table}

\end{document}